\documentclass[twocolumn,epjc3]{svjour3}
\usepackage{mathrsfs,amsmath,latexsym,amssymb,amsfonts,epsfig,cancel,enumerate,graphicx,txfonts,subfigure,cite,graphics,diagbox}
\usepackage{multirow}
\usepackage{booktabs}
\usepackage{xcolor}
\usepackage[colorlinks,linkcolor=blue,citecolor=blue,urlcolor=blue,plainpages=false,pdfstartview=FitH]{hyperref}

\usepackage{url}
\allowdisplaybreaks
\newcommand{\GZU}{School of Physics, Guizhou University, Guiyang 550025, China}
\usepackage{etoolbox}
\usepackage{orcidlink}
\usepackage[english]{babel}
\usepackage{lipsum}
\journalname{Eur. Phys. J. C}

\begin{document}

\title{Quasinormal modes of charged covariant effective black holes with a cosmological constant}

\author{Zhongzhinan Dong
\thanksref{addr1,e1}
\orcidlink{0009-0000-5961-4398}
\and Jinsong Yang
\thanksref{addr1,e2}
\orcidlink{0000-0003-4051-2767}
}

\institute{\GZU\label{addr1}}

\thankstext{e1}{e-mail: gs.zzndong25@gzu.edu.cn}
\thankstext{e2}{e-mail: jsyang@gzu.edu.cn (corresponding author)}

\date{Received: date / Accepted: date}
\maketitle

\begin{abstract}
In this paper, we study the quasinormal modes of two covariant effective black holes characterized by the quantum parameter $\zeta$, charge $Q$, and cosmological constant $\Lambda$, under the scalar perturbation. By employing the pseudo-spectral method, we numerically calculate the quasinormal frequencies and analyze the influence of $\zeta$ on the spectra with respect to $Q$. Our results demonstrate that while the quantum parameter $\zeta$ significantly modifies the quasinormal frequency spectrum, the non-monotonic behavior and overtone outbursts persist. Notably, the impact of quantum gravity on the overtone outbursts is not merely limited to enhancement or suppression; instead, it introduces additional spectral features. Furthermore, a comprehensive analysis of the full quasinormal mode spectrum reveals rich interactions between complex and purely imaginary modes, including damping-rate crossings and splitting-merging behavior. These phenomena typically accompany overtone outbursts in near-extremal regimes, suggesting a potential connection between mode interactions and outbursts. This work emphasizes the necessity of analyzing the full quasinormal frequency spectrum rather than focusing solely on fundamental modes, and provides novel insights into its underlying spectral structures.
\end{abstract}

\maketitle

\section{Introduction}

General relativity, as a successful theory of gravitation, has achieved numerous successes in both theoretical derivations and observational tests~\cite{Wald:1984rg,Liang:2023ahd}. According to the predictions of general relativity, the center of a black hole contains a singularity, where the spacetime curvature diverges and the classical theory ceases to be valid~\cite{Penrose:1964wq,Janis:1968zz,Hawking:1966sx,Joshi:2011rlc}. So far, the problem of singularities in black holes remains one of the most challenging issues in gravitational theory. To address the theoretical difficulties posed by singularities, researchers have explored new directions. Among them, quantum theories of gravity are widely expected to provide a fundamental resolution to this problem~\cite{Thiemann:2007pyv}.

Among the various quantum-corrected models addressing the problem of black hole singularities, phenomenologically constructed models often lack a rigorous foundation in a consistent quantum theory of gravity. In contrast, loop quantum gravity (LQG), with its non-perturbative formulation and background independence, has emerged as a promising candidate, achieving significant breakthroughs~\cite{Thiemann:2007pyv,Rovelli:2004tv,Han:2005km,Giesel:2012ws,Yang:2016kia,Yang:2015zda,Yang:2021den}. As a symmetry-reduced model of LQG, loop quantum cosmology (LQC) has led to notable advances~\cite{Bojowald:2003mc,Tsujikawa:2003vr,Bojowald:2001xe,Bojowald:2005epg,Ashtekar:2003hd,Singh:2009mz,Ashtekar:2006uz}. Similarly, in the study of loop quantum black holes (LQBHs), numerous models have been proposed in recent years, providing new perspectives on resolving black hole singularities~\cite{Zhang:2023yps,Ashtekar:2018lag,Zhang:2021wex,Lin:2024flv,Alonso-Bardaji:2022ear,Gambini:2013ooa,Peltola:2009jm}. However, most LQBH models continue to face theoretical challenges, notably the breaking of diffeomorphism invariance~\cite{Bojowald:2020dkb,Bojowald:2020unm,Bojowald:2015zha}. One approach constructs an effective Hamiltonian constraint incorporating quantum gravity effects while preserving covariance to address the covariance challenge~\cite{Alonso-Bardaji:2023vtl,Zhang:2024khj,Zhang:2024ney,Belfaqih:2024vfk}. Moreover, a fundamental connection between general covariance and Birkhoff's theorem has been revealed in~\cite{Zhang:2025ccx}. Evolving from its vacuum formulation, the approach was extended to include the dust coupling~\cite{Zhang:2024ney} and then generalized to the electrovacuum case with a cosmological constant~\cite{Yang:2025ufs}.

These developments have yielded multiple quantum-corrected solutions, which continue to attract significant attention. A number of studies have focused on the phenomenological and observational properties of covariant effective quantum black holes, including quasinormal modes (QNMs)~\cite{Lin:2024beb,Konoplya:2024lch,Skvortsova:2024msa,Bolokhov:2024bke,Konoplya:2025hgp,Lutfuoglu:2025hwh,Liu:2025iby}, gravitational lensing~\cite{Liu:2024wal,Wang:2024iwt}, and black hole shadows~\cite{Konoplya:2024lch,Liu:2024soc,Chen:2025ifv}, among other related studies~\cite{Li:2026ogy,Du:2024ujg,Umarov:2025wzm,Chen:2025aqh,Yang:2026syx,Chen:2026oxr}. Moreover, several fundamental properties of these models have also been studied~\cite{Sahlmann:2025fde,Lin:2024beb,Motaharfar:2025ihv,Wang:2025alf,Liu:2024pui}. In particular, Ref.~\cite{Liu:2024pui} demonstrated that quantum corrections can induce a repulsive gravitational effect near the Planck scale, in sharp contrast to the purely attractive nature of classical general relativity. More recently, these studies have been further extended to quantum-information aspects~\cite{Du:2025kcx,Liu:2025hcx}. Besides these theoretical investigations, increasing efforts have also been devoted to confronting covariant effective quantum gravity with astrophysical observations, thereby placing constraints on the quantum parameters~\cite{Ai:2025myf,Yao:2026pjz,Calza:2025mwn}.

Among various observational probes, QNMs have attracted particular attention, serving as a bridge between general relativity and quantum gravity, offering a viable approach to test quantum-corrected models. Gravitational waves, which directly mirror the spacetime geometry of black holes, stand as the most direct tool for probing their properties. During the ringdown phase, black holes emit gravitational waves with distinct quasinormal frequencies (QNFs). These frequencies encode information about decay timescales and damped oscillations, thereby revealing internal structure and stability of the black holes~\cite{Berti:2009kk,Konoplya:2011qq,Kokkotas:1999bd,Berti:2004md}. Notably, the real and imaginary parts of QNFs correspond uniquely to fundamental parameters of black hole, enabling precise inference of these parameters and rigorous testing of gravitational theories~\cite{Echeverria:1989hg,Berti:2005ys,Berti:2007zu,LIGOScientific:2025rid,Barausse:2014tra,Cardoso:2016oxy}. Furthermore, quantum gravity effects can further alter the geometric structure of black holes, leaving unique signatures in the QNFs and opening up a promising avenue for investigating quantum gravity~\cite{Berti:2005ys,Berti:2018vdi,Fu:2023drp,Gong:2023ghh}.

In asymptotically de Sitter (dS) spacetimes, the late-time decay of perturbations differs drastically from that in asymptotically flat spacetimes. Unlike the power-law decay exhibited by the latter, the late-time tail in dS spacetime exhibits an exponential decay. Further research has revealed that this exponential decay stems from purely imaginary modes, whose real part vanishes (signifying a non-oscillatory decay process). These modes satisfy the boundary conditions for QNMs in dS spacetimes and thus form part of the QNFs~\cite{Konoplya:2022xid,Konoplya:2024ptj}. Notably, the existence of such purely imaginary modes is closely related to the Strong Cosmic Censorship Conjecture (SCCC). The rapid exponential decay they induce further suppresses the blueshift effect at the Cauchy horizon, which may ultimately lead to a violation of SCCC~\cite{Cardoso:2017soq,Cardoso:2018nvb,Destounis:2018qnb,Liu:2019lon,Shao:2023qlt}.

Beyond the purely imaginary modes, overtones in black hole ringdown signals also play a crucial role. They provide more detailed insights into the properties of black holes~\cite{Giesler:2019uxc,Forteza:2021wfq,Oshita:2021iyn,Oshita:2022pkc}. These findings challenge the conventional view that black hole signals are dominated by the fundamental modes and further motivate in-depth investigations of overtones. Among these studies, Ref.~\cite{Konoplya:2022pbc} revealed that even minor modifications near the horizon can lead to significant changes in the overtone spectrum. This phenomenon, known as the overtone outburst, has also been vividly described as the ``sound of the event horizon". Subsequently, this phenomenon has been extensively investigated in various modified black hole models. The phenomenon is particularly prominent at low angular quantum numbers $l$, often exhibiting non-monotonic or even oscillatory behavior in QNF spectrum. As $l$ increases, such behavior becomes suppressed, but may reappear at sufficiently high overtones~\cite{Gong:2023ghh,Fu:2023drp,Zhang:2024nny,Zhu:2024wic,Bolokhov:2023ozp,Tang:2024txx}. Further studies have shown that the sensitivity of overtone outbursts to near-horizon structures makes them an effective tool for distinguishing different black hole models~\cite{Tang:2024txx,Zhang:2024nny,Fu:2023drp,Saka:2025xxl}; meanwhile, related studies have further explored their connections with physical processes such as Hawking evaporation and grey-body factors~\cite{Bolokhov:2023ozp,Konoplya:2023ppx,Lutfuoglu:2025ljm}.

Recent studies have also explored overtone properties of covariant effective quantum black holes. In particular, Ref.~\cite{Konoplya:2024lch} systematically investigated scalar, electromagnetic, and gravitational perturbations of the covariant effective quantum black holes proposed in~\cite{Zhang:2024khj}. The study found that the fundamental mode depends relatively mildly on the quantum parameter, whereas the first several overtones can exhibit pronounced deviations from the Schwarzschild limit, with the real parts of the QNFs rapidly approaching zero as the quantum parameter increases. These results demonstrated that higher overtones are considerably more sensitive to the near-horizon quantum corrections than the fundamental mode, highlighting their potential as sensitive probes of quantum-gravity effects. This analysis was subsequently extended in Ref.~\cite{Bolokhov:2024bke} to massive scalar perturbations, which revealed the existence of long-lived modes in the QNF spectrum of covariant effective quantum black holes. Since the QNFs of Reissner-Nordstr\"om black holes also exhibit anomalous overtone behaviors near the extremal limit~\cite{Jing:2008an}, the combined effects of charge and quantum corrections may further modify the QNF spectrum and lead to distinctive spectral features. However, previous studies have mostly investigated the effects of charge or quantum corrections on QNFs separately, while the combined effects of charge and quantum corrections on QNMs have received considerably less attention, particularly for asymptotically dS spacetimes.

Motivated by this gap, building upon the analyses of Refs.~\cite{Bolokhov:2024bke,Konoplya:2024lch}, in the present work we investigate two charged covariant effective black holes with a cosmological constant proposed in Ref.~\cite{Yang:2025ufs} (IIIC and IIID, case 2). To focus on the effects of black hole parameters themselves, we consider massless scalar perturbations and employ the pseudo-spectral method (PSM) to calculate and analyze the corresponding QNFs. We aim to analyze how the quantum parameter $\zeta$ affects the spectral characteristics, thereby revealing the potential interplay between quantum gravity effects and charge. Of particular significance, the purely imaginary modes appearing in these spectra encode richer physical information. A systematic investigation of these features is expected to shed new light on the properties of QNMs of quantum-corrected black holes, and provide a fresh perspective for testing candidate theories of quantum gravity.

This paper is structured as follows. Section~\ref{section2} introduces the covariant quantum-corrected black hole solutions and discusses the relevant parameter constraints. In Sect.~\ref{section3}, we derive the massless scalar perturbation equation and analyze the properties of the corresponding effective potential. Section~\ref{section4} is dedicated to the properties of the QNM spectra. The conclusions and discussions are presented in Sect.~\ref{section5}. Further technical details concerning the PSM implementation and the cosmological constant choice are provided in \ref{app:A} and \ref{app:B}.

\section{Covariant effective spacetimes}
\label{section2}

In~\cite{Zhang:2024khj, Zhang:2024ney}, the authors proposed an algebraic framework to address the covariance issues in the construction of effective quantum-corrected black holes. In~\cite{Yang:2025ufs}, this framework was further extended to the electro-vacuum case with a cosmological constant, leading to the derivation of several quantum-corrected black hole solutions. In this work, we focus on the first two quantum-corrected black hole spacetimes with the following line elements~\cite{Yang:2025ufs}:
\begin{equation}\label{original_metric}
	\mathrm{d}s^2=-f(r)\mathrm{d}t^2+\left[g(r)f(r)\right]^{-1}\mathrm{d}r^2+r^2\mathrm{d}\Omega^2\,,
\end{equation}
where $\mathrm{d}\Omega^2=\mathrm{d}\theta^2+\sin^2\theta \mathrm{d}\phi^2$. In Solution 1, the metric functions read:
\begin{equation}\label{metric1}
	\begin{aligned}
		f_0(r) &= 1-\frac{2M}{r}+\frac{Q^2}{r^2}-\frac{\Lambda r^2}{3}, \\
		f(r) &= f_0(r)\left[1+\frac{\zeta^2}{r^2} f_0(r)\right], \\
		g(r) &= 1\,.
	\end{aligned}
\end{equation}
For Solution 2, the metric functions take the form:
\begin{equation}\label{metric2}
	\begin{aligned}
		f(r)&=1-\frac{2M}{r}+\frac{Q^2}{r^2}-\frac{\Lambda r^2}{3}\,,\\
		g(r)&=1+\frac{\zeta^2}{r^2}f(r)\,.
	\end{aligned}
\end{equation}
Here, $M$ denotes the black hole mass, $\Lambda$ is the cosmological constant, $Q$ is the charge, and $\zeta$ is a quantum parameter. When $\zeta=0$, both solutions reduce to the Reissner-Nordstr\"{o}m-de Sitter (RN-dS) black hole solution. In asymptotically dS spacetimes, the study of QNMs is restricted to the region between the event horizon $r_+$ and the cosmological horizon $r_c$. For practical purposes, especially when computing QNFs, it is often convenient to express $M$ and $\Lambda$ in terms of the horizons.

The horizon structure of both solutions is determined by the parameters. Specifically, the charge $Q$ affects the positions of the event horizon $r_{+}$ and the cosmological horizon $r_{c}$, as well as the structure of the inner horizon(s), which may consist of one or more horizons depending on the value of charge $Q$ and the quantum parameter $\zeta$. The cosmological constant $\Lambda$ also influences the overall horizon structure. By contrast, when the quantum parameter $\zeta$ is below a certain critical value, the event horizon $r_+$ and the cosmological horizon $r_c$ are independent of $\zeta$. However, when $\zeta$ exceeds this value, an additional outer horizon emerges outside the original cosmological horizon, leading to a change in the spacetime structure. Since QNMs are studied in the region between the event horizon and the outermost boundary, such a structural change modifies the domain of the problem, and the corresponding spectra may not be continuously connected across this transition. Therefore, in this work we focus on the regime before this structural change occurs, i.e. $r_+$ and $r_c$ are independent of $\zeta$.

Under this condition, solving $f(r_+)=0$ and $f(r_c)=0$ yields expressions for the mass $M$ and cosmological constant $\Lambda$ that depend only on $r_+$, $r_c$, and the charge $Q$. They are then given by:
\begin{equation}
	\begin{aligned}
		M&=\frac{\left(r_++r_c\right)\left(Q^2\left(r_+^2+r_c^2\right)+r_+^2r_c^2\right)}{2\left(r_+^3r_c+r_+^2r_c^2+r_c^3r_+\right)}\,,\\
		\Lambda&=\frac{3r_+r_c-3Q^2}{r_+^3r_c+r_+^2r_c^2+r_c^3r_+}\,.
	\end{aligned}
\end{equation}
For convenience in the following analysis, we normalize the black hole mass $M$ as a reference scale so that all relevant parameters can be expressed in dimensionless form. For example, $\hat{\Lambda} \rightarrow \Lambda M^2$, $\hat{Q} \rightarrow Q/M$, and $\hat{\zeta} \rightarrow \zeta/M$. In the following, for notational simplicity, we omit the hat symbols and continue to denote the corresponding dimensionless quantities by $\Lambda$, $Q$, and $\zeta$. This convention does not affect the physical interpretation of the results, which remain expressed in terms of dimensionless combinations of parameters.

\begin{figure}[htbp]
	\centering
	\includegraphics[width=3in, height=5.5in, keepaspectratio]{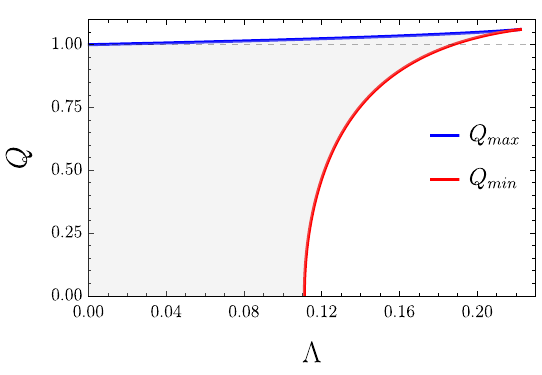}

	\caption{The maximum and minimum charge values, $Q_{\max}$ and $Q_{\min}$, as functions of the cosmological constant $\Lambda$ for the RN-dS black hole. The upper bound $Q_{\max}$ corresponds to the Nariai limit.}
	\label{parameter}
\end{figure}

Previous analyses show that, in the regime without the emergence of an additional outer horizon, the quantum parameter $\zeta$ affects only the inner horizons. Therefore, the allowed ranges of $Q$ and $\Lambda$ can be determined from the $\zeta=0$ case, i.e., the RN-dS black hole. The corresponding $\Lambda$-$Q$ diagram is shown in Fig.~\ref{parameter}. For a fixed $\Lambda$, the admissible charge range $Q_{\min} < Q < Q_{\max}$ is obtained from the horizon degeneracy conditions $f(r)=0$ and $\mathrm{d}f(r)/\mathrm{d}r=0$, where $Q_{\min}$ corresponds to the inner and event horizon merger and $Q_{\max}$ to the Nariai limit with coincident event and cosmological horizons. Outside this region, the spacetime no longer describes a static RN-dS black hole with three horizons. In this work, we restrict our analysis to the parameter range $Q \in (0,1)$ and $\Lambda \in (0,1/9)$, ensuring non-degenerate horizon configurations.

\begin{figure}[htbp]
	\centering
	\includegraphics[width=3in, height=5.5in, keepaspectratio]{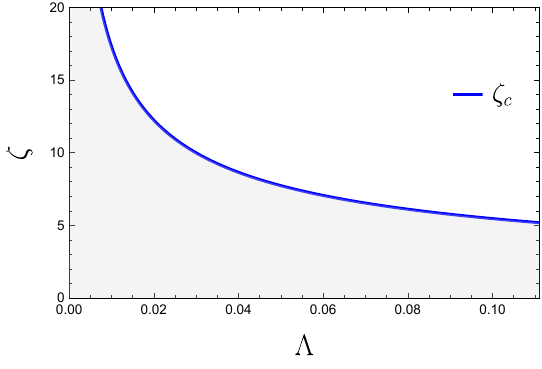}

	\caption{The critical value $\zeta_c$ as a function of the cosmological constant $\Lambda$, separating the parameter region without an additional outer horizon from the region where an extra outer horizon appears.}
	\label{parament2}
\end{figure}

The emergence of an additional outer horizon is controlled by a critical value $\zeta_c$, shown in Fig.~\ref{parament2}, which depends on $\Lambda$ and decreases monotonically with it, reaching its minimum at $\Lambda=1/9$, while remaining independent of $Q$. Meanwhile, we can further consider the restriction $0 < \zeta \lesssim 7.09$, set by the minimum black hole mass in effective LQG~\cite{Lin:2024beb}. In the parameter range considered in this work, both conditions $\zeta < \zeta_c$ and $0 < \zeta \lesssim 7.09$ are satisfied.

\section{Scalar field over the black hole}
\label{section3}

In this section, we investigate how those effective quantum-corrected black holes respond to perturbations from a massless scalar field $\Phi$. The scalar field's dynamics is governed by the Klein-Gordon equation:
\begin{equation}
	\frac{1}{\sqrt{-g}}\partial_\mu(\sqrt{-g}g^{\mu\nu}\partial_\nu\Phi)=0\,,
\end{equation}
where $g_{\mu\nu}$ is the background metric. We will analyze the eigenvalue problem associated with the dynamics of a scalar field in the frequency domain. After the separation of variables, the aforementioned equation can be rewritten into the Schr\"odinger-like form:
\begin{equation}\label{perturbationeq1}
	\frac{\partial^2\Psi}{\partial r^2_*}+\left(\omega^2-V_{\mathrm{eff}}\right)\Psi=0\,.
\end{equation}
Here, $r_*$ is the tortoise coordinate, defined by the relation $\mathrm{d}r_*=\mathrm{d}r/\left(f(r)\sqrt{g(r)}\right)$, $\Psi$ is the radial function arising from the separation of variables, and the effective potential $V_{\mathrm{eff}}$ is given by:
\begin{equation}
		V_{\mathrm{eff}}=f(r)\left[\frac{l(l+1)}{r^2}+\frac{\sqrt{g(r)}}{r}\frac{\mathrm{d}}{\mathrm{d}r}\left(f(r)\sqrt{g(r)}\right)\right]\,.
\end{equation}

The effective potential consists of two main parts. The first part, associated with the angular quantum number $l$, is known as the centrifugal potential. It prevents the wave from approaching the center and forms a barrier in regions farther away. The second part is the gravitational potential, determined by the spacetime geometry. To further analyze how different parameters affect the effective potential, we introduce the quantity
\begin{equation}\label{ratio}
	\frac{{\rm d}V_{\mathrm{eff}}/{\rm d}x}{V_{\mathrm{eff}}|_{x=0}}\,,
\end{equation}
which characterizes the relative variation of $V_{\mathrm{eff}}$ with respect to the parameter $x$. Here, $x$ denotes the parameter under consideration, such as the quantum parameter $\zeta$ or the charge $Q$.

For quantum-corrected black holes, when the quantum parameter $\zeta$ is chosen as $x$, the value defined in Eq.~\eqref{ratio} generally decreases with increasing angular quantum number $l$, suggesting that the effects of $l$ tend to suppress the contributions from quantum gravity. Moreover, this suppression can be further reflected in the QNFs, and has been observed in previous studies of QNMs in quantum-corrected black hole spacetimes~\cite{Gong:2023ghh,Fu:2023drp,Zhang:2024nny}. In particular, in Solution 2, the quantum parameter $\zeta$ appears only in the gravitational potential. As a result, the quantum gravity effects are more strongly suppressed by the angular quantum number $l$ in this model, leading to more pronounced differences between the cases $l=0$ and $l=1$. Therefore, to analyze the impact of quantum gravity effects, we focus mainly on the cases $l=0$ and $l=1$.
\begin{figure*}[htbp]
	\centering
	\subfigure[$l=0,~Q=0.5$]{\includegraphics[width=0.24\textwidth,keepaspectratio]{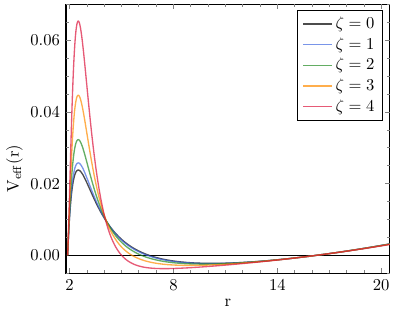}}
	\hspace{-0.3em}
	\subfigure[$l=1,~Q=0.5$]{\includegraphics[width=0.24\textwidth,keepaspectratio]{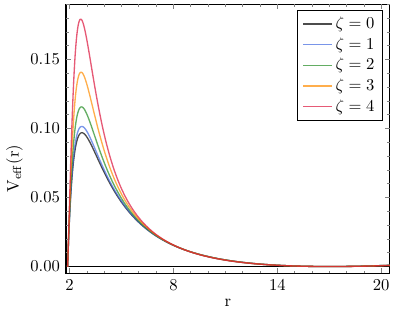}}
	\hspace{-0.3em}
	\subfigure[$l=0,~\zeta=2$]{\includegraphics[width=0.24\textwidth,keepaspectratio]{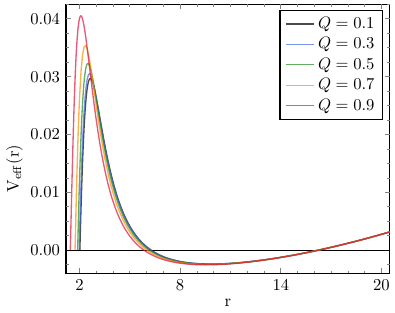}}
	\hspace{-0.3em}
	\subfigure[$l=1,~\zeta=2$]{\includegraphics[width=0.24\textwidth,keepaspectratio]{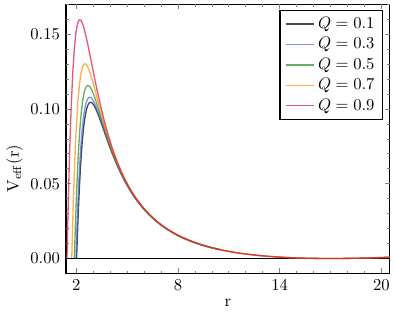}}
	\hspace{-0.3em}

	\caption{Effective potential $V_{\mathrm{eff}}(r)$ of Solution 1 for $\Lambda = 0.01$ with $l = 0$ and $l = 1$. The left two panels, (a) and (b), correspond to fixed $Q$ and varying $\zeta$, whereas the right two panels, (c) and (d), show fixed $\zeta = 2$ with varying $Q$.}
	\label{metric1_veff}
\end{figure*}
\begin{figure*}[htbp]
	\centering
	\subfigure[$l=0,~Q=0.5$]{\includegraphics[width=0.24\textwidth,keepaspectratio]{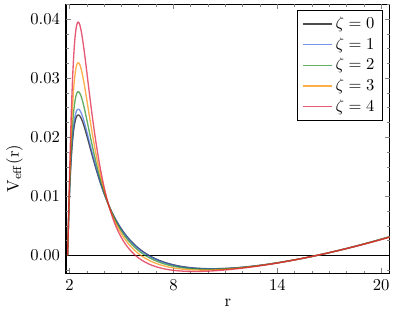}}
	\hspace{-0.3em}
	\subfigure[$l=1,~Q=0.5$]{\includegraphics[width=0.24\textwidth,keepaspectratio]{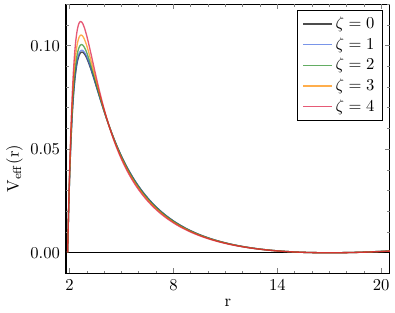}}
	\hspace{-0.3em}
	\subfigure[$l=0,~\zeta=2$]{\includegraphics[width=0.24\textwidth,keepaspectratio]{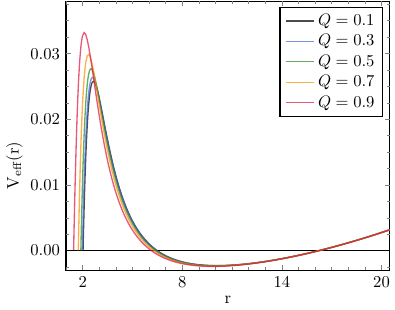}}
	\hspace{-0.3em}
	\subfigure[$l=1,~\zeta=2$]{\includegraphics[width=0.24\textwidth,keepaspectratio]{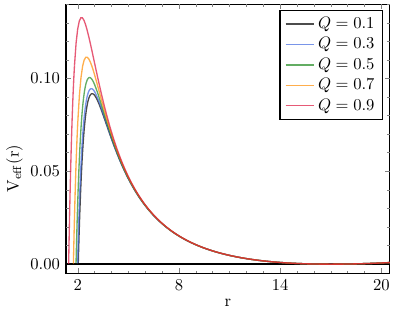}}
	\hspace{-0.3em}

	\caption{Effective potential $V_{\mathrm{eff}}(r)$ of Solution 2 for $\Lambda = 0.01$ with $l = 0$ and $l = 1$. The left two panels, (a) and (b), correspond to fixed $Q$ and varying $\zeta$, whereas the right two panels, (c) and (d), show the case of fixed $\zeta = 2$ with varying $Q$.}
	\label{metric2_veff}
\end{figure*}

We first provide a qualitative examination of the effective potentials $V_{\mathrm{eff}}$, focusing on the relationship between its structural features and the characteristics of the QNFs. As shown in Figs.~\ref{metric1_veff} and \ref{metric2_veff}, for a nonzero cosmological constant, the effective potentials develop local negative regions after the peak, before eventually returning to positive values and increasing monotonically at large radial distances. These figures indicate that for small values of the cosmological constant, the two models display similar parameter dependence: at fixed $l$, increasing $\zeta$ or $Q$ enlarges the negative potential region. Moreover, at fixed $\zeta$ and $Q$, the case $l=0$ shows a wider negative region than $l=1$.

A deeper or wider negative well implies a more complicated behavior of the wave function in that region. This increased complexity is generally associated with a stronger sensitivity of the perturbations to the background geometry and may indicate a greater susceptibility to black hole instability in more extreme scenarios~\cite{Guo:2019tjy}. A further comparison shows that for $l=0$, with all other parameters fixed, Solution 1 exhibits a larger negative region in the effective potential than Solution 2, which may lead to increased numerical difficulty in computing higher-order overtones. Fortunately, within the parameter range explored in this work, all QNFs satisfy $\mathrm{Im}(\omega)<0$, and the system remains stable.

Next, we perform a detailed analysis of the effective potential. First, we consider the case of fixed charge $Q$ and cosmological constant $\Lambda$ (see Figs.~\ref{metric1_veff} and \ref{metric2_veff}, left two panels). The peak of the effective potential increases with the quantum parameter $\zeta$. Compared with Solution 1, Solution 2 exhibits a more pronounced dependence on the angular quantum number $l$. For fixed $\zeta$ and $\Lambda$ (see Figs.~\ref{metric1_veff} and \ref{metric2_veff}, right two panels), the two solutions exhibit nearly identical effective potentials for different values of $Q$, with only minor differences in magnitude. This indicates that $Q$ acts as a subleading parameter in the effective potential structure, and does not significantly affect the qualitative difference between the two solutions when $\zeta$ is fixed. This observation is consistent with Eq.~\eqref{ratio}, which yields significantly smaller variations for $x=Q$ than for $x=\zeta$.

In view of this, in the subsequent analysis we study the dependence of QNFs on $Q$ for different fixed values of $\zeta$. Since the influence of $Q$ on the effective potential is comparatively weaker, scanning $Q$ at different $\zeta$ allows the effects of quantum corrections on the spectra to be distinguished more clearly.

\section{Numerical results}
\label{section4}

In this section, we examine the numerical results for the QNFs of the two covariant quantum-corrected black holes defined in Eqs.~\eqref{metric1} and \eqref{metric2}. We will analyze how the quantum parameter $\zeta$ manifests itself in the QNFs and highlight the differences between the two solutions. The QNFs are obtained by solving the eigenvalue problem subject to the boundary conditions of a purely outgoing wave at the cosmological horizon and a purely ingoing wave at the event horizon, namely,
\begin{equation}
	\begin{aligned}\label{boundary_condition}
		\Psi&\sim e^{+i\omega r_*}\,,~~~~~r_*\rightarrow+\infty\,,\\
		\Psi&\sim e^{-i\omega r_*}\,,~~~~~r_*\rightarrow-\infty\,.
	\end{aligned}
\end{equation}
The above boundary conditions describe the black hole’s response to transient perturbations once the source has ceased to act~\cite{Konoplya:2011qq,Berti:2009kk,Kokkotas:1999bd}.

Regarding the computation of QNFs, several numerical methods have been developed, such as the WKB method~\cite{Ferrari:1984zz,Schutz:1985km,Iyer:1986np,Iyer:1986nq}, the asymptotic iteration method (AIM)~\cite{Ciftci:2005xn,Cho:2009cj,Cho:2011sf}, the continued fraction method (CFM)~\cite{Leaver:1985ax,Leaver:1986vnb}, and the PSM~\cite{Boyd2000,Jansen:2017oag}, among others.

The WKB method is one of the most widely used approaches for calculating quasinormal modes because of its simplicity and good accuracy for low-lying modes. Recent developments combining very high-order WKB expansions with the Cesàro convergence technique have further extended its applicability to higher overtones~\cite{Konoplya:2026rjh}. However, since the present work also involves purely imaginary modes, which cannot be captured within the WKB framework, this method is not suitable for our analysis. Although the CFM can in principle determine such modes, it often requires accurate initial guesses and higher iterations. By contrast, the PSM formulates the problem as a matrix eigenvalue problem and can robustly compute both complex and purely imaginary QNMs. This ensures reliable results and offers clear advantages for studying purely imaginary modes in dS (or AdS) spacetimes. We therefore employ the PSM throughout this work.

In the PSM, Eq.~\eqref{perturbationeq1} is discretized on Chebyshev grids, transforming the perturbation equation into the matrix eigenvalue problem
\begin{equation}\label{matrix_eq}
	\begin{aligned}
		\mathcal{M}(\omega)\delta\psi&=(M_0+\omega M_1+\omega^2M_2+\cdots+\omega^PM_P)\delta\psi\\
		&=0\,.
	\end{aligned}
\end{equation}

To implement this method, the computational domain needs to be mapped onto the interval $(0,1)$, which is typically facilitated by the coordinate transformation $u = 1/r$. Moreover, to enhance the numerical stability and efficiency of solving this matrix eigenvalue problem, it is advantageous to employ the Eddington-Finkelstein coordinate system. In this coordinate system, the frequency term $\omega$ in the final matrix equation Eq.~\eqref{matrix_eq} exhibits a lower power order compared to other coordinate choices. This reduction simplifies the structure of the matrix equation, thereby enabling a more robust and accurate computation of the eigenvalues.

Considering $v=t+r_*$ and $u = 1/r$, the original metric \eqref{original_metric} takes the form:
\begin{equation}\label{ef_metric}
	\mathrm{d}s_{EF}^2=-f(u)\mathrm{d}v^2-\frac{2\mathrm{d}v\mathrm{d}u}{u^2\sqrt{g(u)}}+\frac{1}{u^2}\mathrm{d}\Omega^2\,.
\end{equation}
Simultaneously, the parameters $M$ and $\Lambda$ become:
\begin{equation}
	\begin{aligned}
		M&=\frac{\left(u_++u_c\right)\left(Q^2\left(u_+^2+u_c^2\right)+1\right)}{2\left(u_+^2+u_+u_c+u_c^2\right)}\,,\\
		\Lambda&=\frac{3u_+^2u_c^2\left(Q^2u_+u_c-1\right)}{u_+^2+u_+u_c+u_c^2}\,.
	\end{aligned}
\end{equation}
Next, using $u=1/r$ and writing the radial function as $\psi(u)$, the wave equation Eq.~\eqref{perturbationeq1} can be transformed into the following form:
\begin{equation}\label{perturbationeq2}
	\begin{aligned}
		&-\left(l\left(l+1\right)+2i\omega\sqrt{g(u)}/u\right)\psi(u)+\left(2i\omega\sqrt{g(u)}+u^2g(u)f^\prime(u)\right.\\
		&\left.+u^2f(u)g^\prime(u)/2\right)\psi^\prime(u)+u^2f(u)g(u)\psi^{\prime\prime}(u)=0\,.
	\end{aligned}
\end{equation}

Combining with the boundary conditions Eqs.~\eqref{boundary_condition}, we can reformulate Eq.~\eqref{perturbationeq2} as a matrix eigenvalue problem, from which the QNFs are obtained. Further details of the implementation are presented in \ref{app:A}.

\begin{figure}[htbp]
		\centering
		\begin{minipage}{0.45\textwidth}
			\includegraphics[width=\textwidth,keepaspectratio]{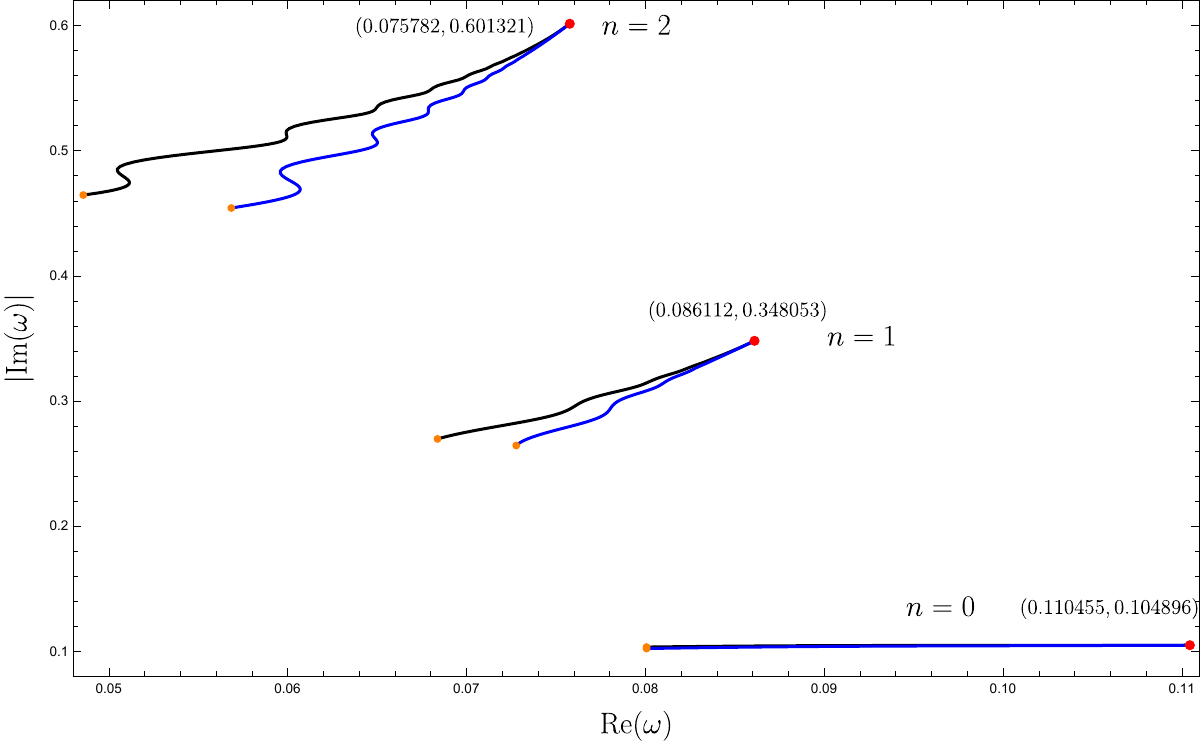}
		\end{minipage}

		\caption{Phase diagrams $\mathrm{Re}(\omega)$-$|\mathrm{Im}(\omega)|$ of the QNFs for two Solutions, showing the fundamental modes and the first few overtone modes for $l=0$. The black and blue lines represent the results for Solution 1 and Solution 2, respectively, as $\epsilon$ varies from $1$ (orange point) to $0$ (red point).}
		\label{r-i}
	\end{figure}

To further assess the reliability of the numerical results, we examined whether the well-established Schwarzschild limit can be recovered. To approach the Schwarzschild limit, we introduce a scaling parameter $\epsilon$ and choose one possible scaling path as follows:
\begin{equation}
	\zeta=\epsilon,\qquad Q=\frac{\epsilon}{2},\qquad \Lambda=0.05\epsilon,
\end{equation}
where $\epsilon\in[0,1]$. As shown in Fig.~\ref{r-i}, the orange and red points represent the QNFs for the initial ($\epsilon=1$) and Schwarzschild ($\epsilon=0$) limits, respectively. The QNFs of both Solutions smoothly recover the Schwarzschild values. A comparison with the corresponding values in Refs.~\cite{Berti:Ringdown,Berti:2005ys,Berti:2009kk} is presented in Table~\ref{Schwarzschild}, with the relative differences in both the real and imaginary parts being below $0.1\%$. These results provide further confidence for the subsequent analysis.

\begin{table}[htbp]
	\centering
		\begin{tabular}{|c|c|c|}
			\hline
			$l$ & $n$ & $\omega$ \\
			\hline
			\multirow{6}{*}{0} & \multirow{2}{*}{0} & 0.110455$-$0.104896\,i \\
			& & 0.1104549399$-$0.1048957201\,i \\
			\cline{3-3}
			& \multirow{2}{*}{1} & 0.086112$-$0.348053\,i \\
			& & 0.0861167750$-$0.3480524669\,i \\
			\cline{3-3}
			& \multirow{2}{*}{2} & 0.075782$-$0.601321\,i \\
			& & 0.0757419355$-$0.6010785900\,i\\
			\hline
		\end{tabular}

	\caption{Comparison of the Schwarzschild QNFs obtained in this work with the corresponding results from Refs.~\cite{Berti:Ringdown,Berti:2005ys,Berti:2009kk}. For each overtone, the upper and lower values represent the results obtained in this work and the corresponding values from the references, respectively.}
	\label{Schwarzschild}
\end{table}

Then, we fix the cosmological constant $\Lambda$ to an appropriate value. Although the cosmological constant $\Lambda$ is extremely small in the observable universe, our numerical analysis shows that increasing $\Lambda$ noticeably alters the effective potential $V_{\mathrm{eff}}$. In addition, it tends to suppress the sensitivity of the QNFs to both charge $Q$ and the quantum parameter $\zeta$. This suppression effect can obscure the intrinsic dependence that we aim to highlight. Therefore, to preserve the essential dS characteristics while allowing a clearer illustration of the dependence of the QNFs on $Q$ and $\zeta$, we fix $\Lambda=0.01$ in all subsequent calculations. Additional results for larger $\Lambda$, demonstrating the suppressive trend while confirming that the qualitative behavior remains unchanged, are provided in \ref{app:B}.

\subsection{Complex quasinormal modes}

\subsubsection{Solution 1}

\begin{figure*}[htbp]
	\centering
	\subfigure[$\zeta=0$ (RN-dS black hole)]{\begin{minipage}{0.33\textwidth}
		\includegraphics[width=\textwidth,keepaspectratio]{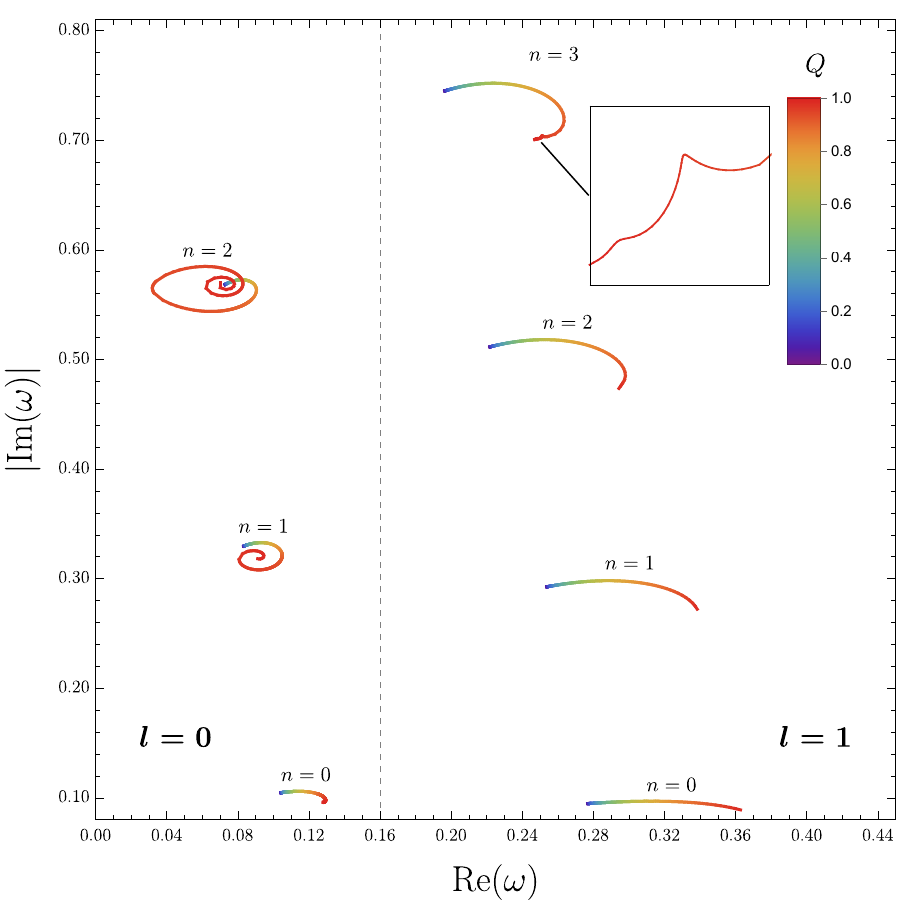}
	\end{minipage}\label{metric1_l=0,1_Re-Im_a}}
	\hspace{-0.55em}
	\subfigure[$\zeta=1$]{\begin{minipage}{0.33\textwidth}
		\includegraphics[width=\textwidth,keepaspectratio]{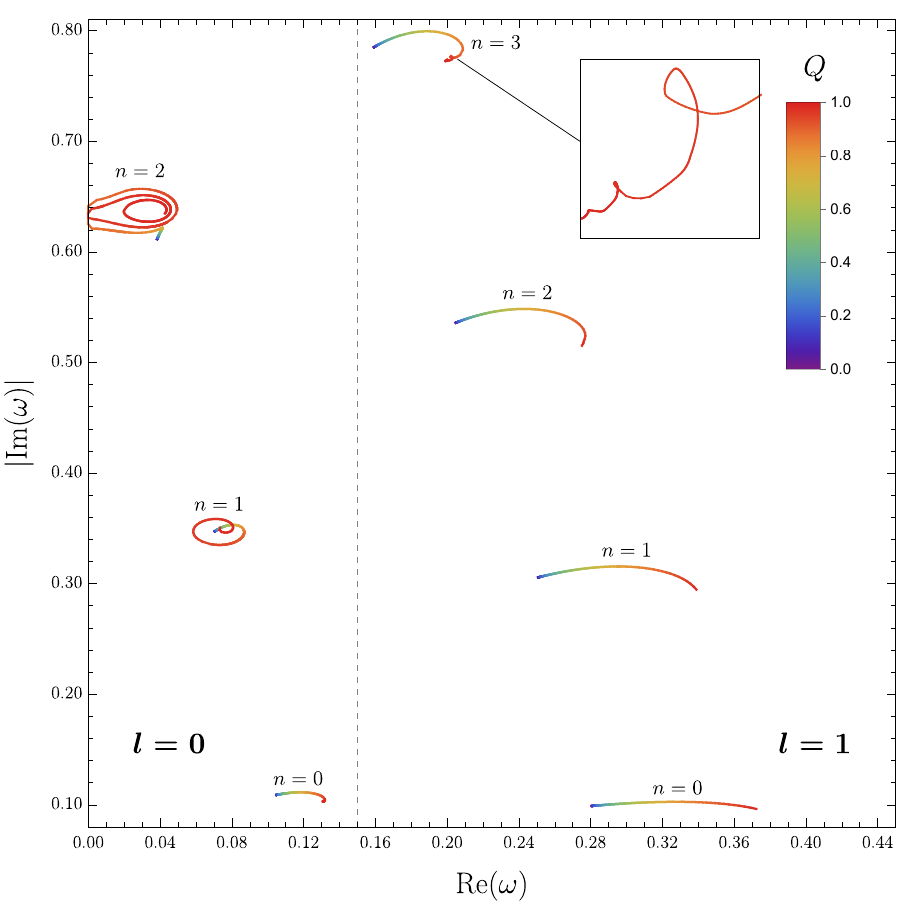}
	\end{minipage}\label{metric1_l=0,1_Re-Im_b}}
	\hspace{-0.55em}
	\subfigure[$\zeta=3$]{\begin{minipage}{0.33\textwidth}
		\includegraphics[width=\textwidth,keepaspectratio]{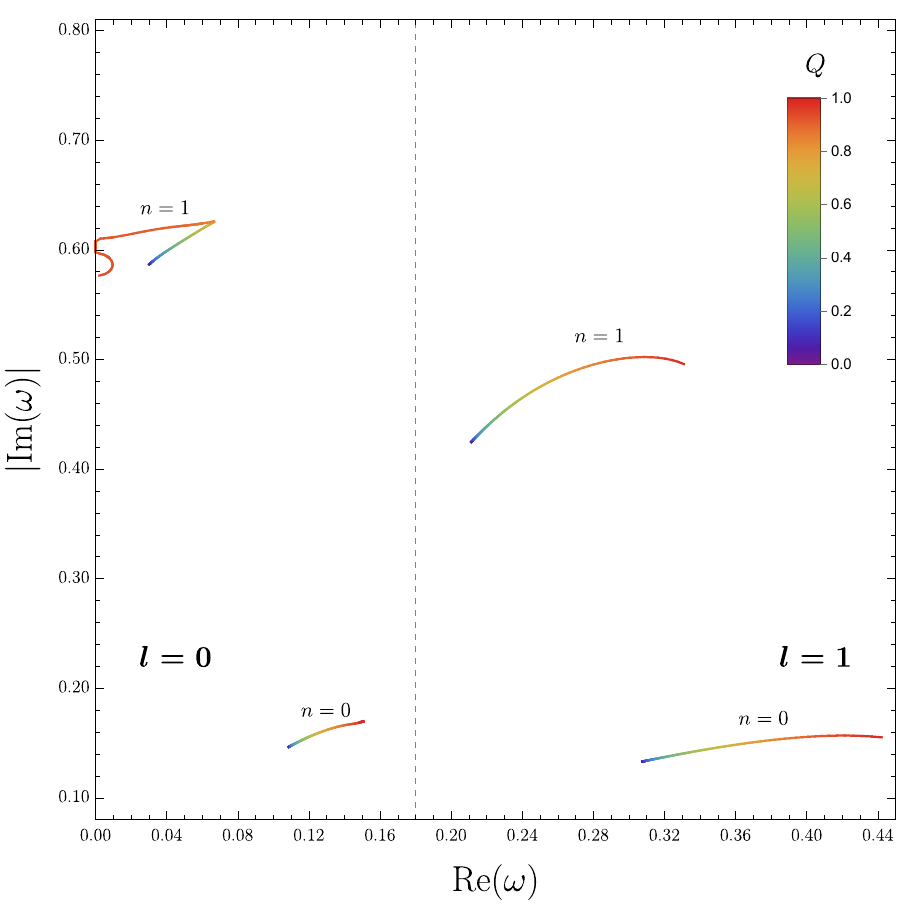}
	\end{minipage}\label{metric1_l=0,1_Re-Im_c}}

	\caption{Phase diagrams $\mathrm{Re}(\omega)$-$|\mathrm{Im}(\omega)|$ of the complex QNFs for Solution 1, which show the fundamental modes and the first few overtone modes. Results are presented for $\Lambda = 0.01$ and $l=0,1$.}
	\label{metric1_l=0,1_Re-Im}
\end{figure*}
\begin{figure}[htbp]
	\centering
	\subfigure[$l=0$]{\begin{minipage}{0.22\textwidth}
			\centering
			\includegraphics[width=\textwidth,keepaspectratio]{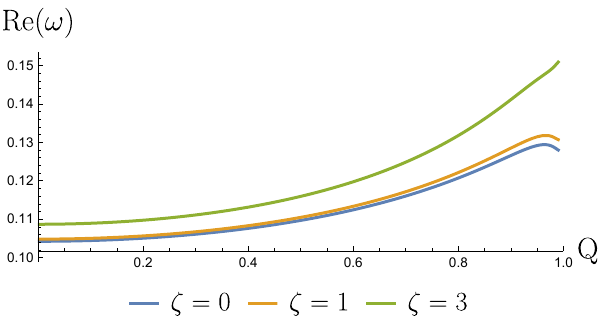}\\[2ex]
			\includegraphics[width=\textwidth,keepaspectratio]{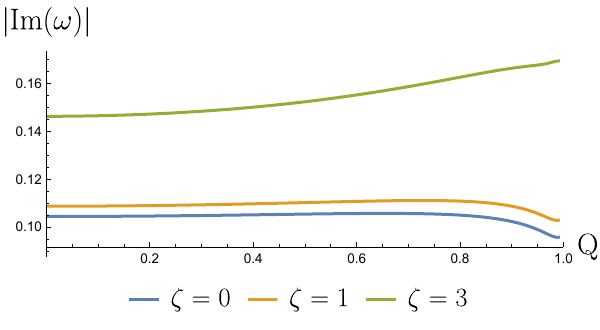}
		\end{minipage}}\label{metric1_Re/Im_fundamental_a}
	\subfigure[$l=1$]{\begin{minipage}{0.22\textwidth}
			\centering
			\includegraphics[width=\textwidth,keepaspectratio]{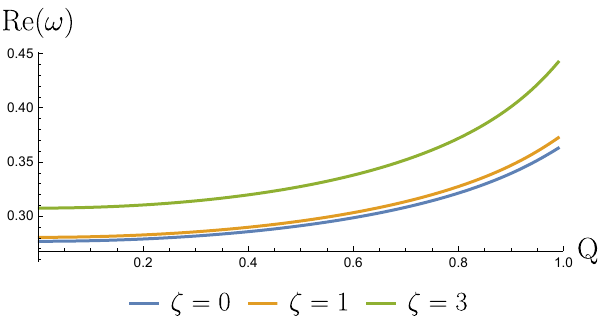}\\[2ex]
			\includegraphics[width=\textwidth,keepaspectratio]{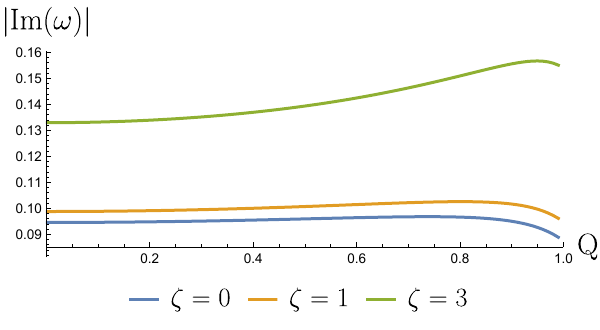}
		\end{minipage}}\label{metric1_Re/Im_fundamental_b}

	\caption{The QNFs of fundamental modes for Solution 1 are presented as a function of the charge $Q$ for $l=0$ and $l=1$.}\label{metric1_Re/Im_fundamental}
\end{figure}
\begin{figure}[htbp]
	\centering
	\subfigure[$l=0,~n=1$]{\begin{minipage}{0.22\textwidth}
		\includegraphics[width=\textwidth,keepaspectratio]{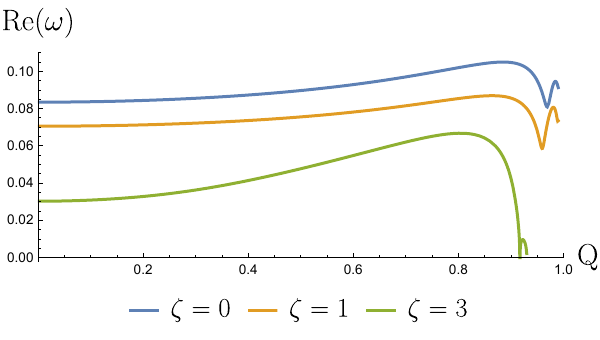}\\[2ex]
		\includegraphics[width=\textwidth,keepaspectratio]{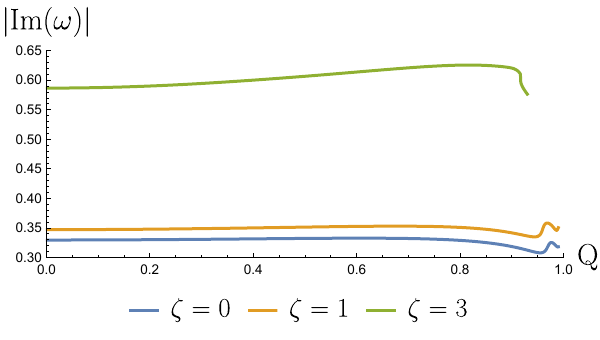}
	\end{minipage}\label{metric1_Re/Im_detail_a}}
	\subfigure[$l=0,~n=2$]{\begin{minipage}{0.22\textwidth}
		\includegraphics[width=\textwidth,keepaspectratio]{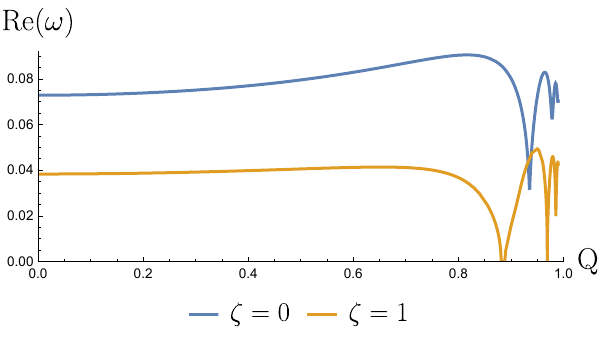}\\[2ex]
		\includegraphics[width=\textwidth,keepaspectratio]{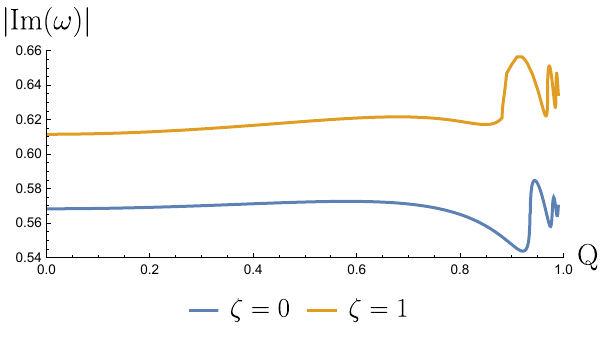}
	\end{minipage}\label{metric1_Re/Im_detail_b}}
	\subfigure[$l=1,~n=1$]{\begin{minipage}{0.22\textwidth}
		\includegraphics[width=\textwidth,keepaspectratio]{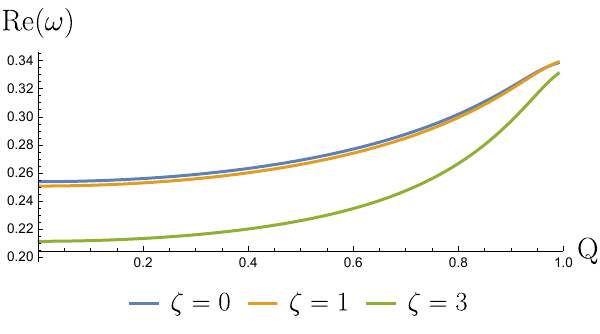}\\[2ex]
		\includegraphics[width=\textwidth,keepaspectratio]{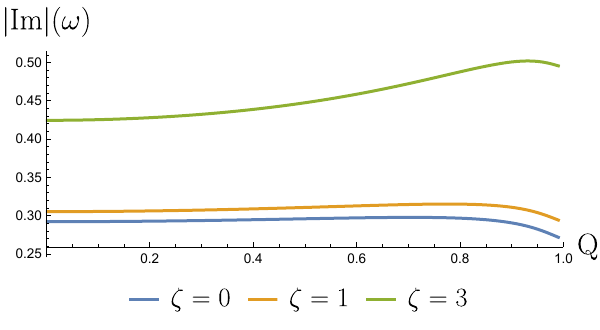}
	\end{minipage}\label{metric1_Re/Im_detail_c}}
	\subfigure[$l=1,~n=3$]{\begin{minipage}{0.22\textwidth}
		\includegraphics[width=\textwidth,keepaspectratio]{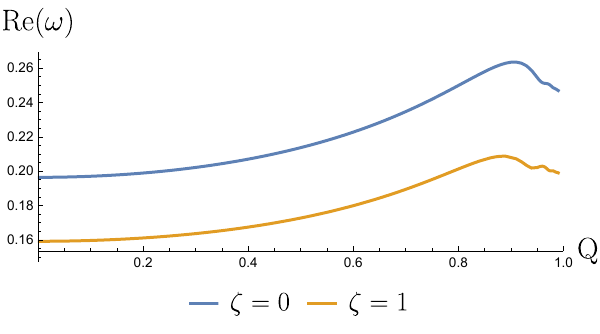}\\[2ex]
		\includegraphics[width=\textwidth,keepaspectratio]{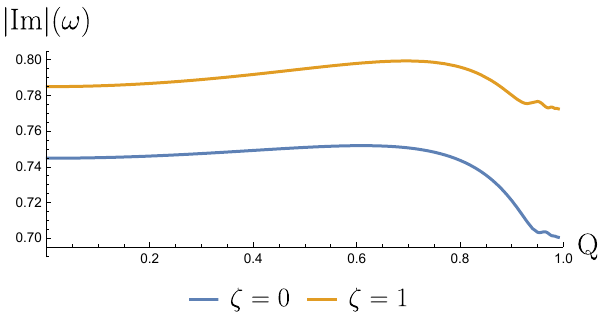}
	\end{minipage}\label{metric1_Re/Im_detail_d}}

	\caption{The QNFs of the first few overtone modes for Solution 1 are presented as a function of the charge $Q$ for $l=0$ and $l=1$.}
	\label{metric1_Re/Im_detail}
\end{figure}
In Fig.~\ref{metric1_l=0,1_Re-Im}, we show the real part $\mathrm{Re}(\omega)$ and the absolute value of the imaginary part $|\mathrm{Im}(\omega)|$ of the QNFs for the first solution, Eq.~\eqref{metric1}, as functions of the charge $Q$ for different values of the quantum parameter $\zeta$. From panels \ref{metric1_l=0,1_Re-Im_a} to \ref{metric1_l=0,1_Re-Im_c}, one can clearly see the influence of $\zeta$ on the QNFs. We first discuss the fundamental complex modes ($n=0$), shown in Fig.~\ref{metric1_Re/Im_fundamental}. For $l=0$ and $\zeta=0$ (i.e., the RN-dS black hole), both $\mathrm{Re}(\omega)$ and $|\mathrm{Im}(\omega)|$ exhibit non-monotonic behavior, initially increasing and then decreasing. This feature persists at $\zeta=1$, while it disappears at $\zeta=3$, where both quantities become monotonic functions of $Q$. This indicates that beyond a certain value of the quantum parameter, the damping-rate no longer exhibits a peak as a function of $Q$.

For $l=1$, the behavior of $\mathrm{Re}(\omega)$ remains monotonic for all values of $\zeta$, while $|\mathrm{Im}(\omega)|$ still exhibits a non-monotonic dependence on $Q$ at $\zeta=0$, which is gradually weakened as $\zeta$ increases. In particular, at $\zeta=3$, the non-monotonic behavior of $|\mathrm{Im}(\omega)|$ is significantly suppressed but does not completely disappear. Compared with the case $l=0$, where the non-monotonic structure is fully removed at $\zeta=3$, this difference suggests that the angular quantum number $l$ further suppresses the influence of quantum gravity effects on the QNFs.

Next, we consider the complex overtones ($n>0$). To clearly illustrate the differences, we show selected overtone results in Fig.~\ref{metric1_Re/Im_detail}. For $\zeta=0$, the overtones exhibit a behavior reminiscent of those in asymptotically flat RN black holes, where the non-monotonic behavior of the fundamental mode is further enhanced, manifesting as a pronounced outburst with oscillations. Specifically, $\mathrm{Re}(\omega)$ first increases with $Q$, then rapidly decreases during the outburst stage before entering the oscillatory regime, whereas $|\mathrm{Im}(\omega)|$ varies smoothly prior to the onset of the outburst. Higher-order overtones exhibit even stronger outburst and oscillatory features.

With the introduction of the quantum parameter, the behaviors of $\mathrm{Re}(\omega)$ and $|\mathrm{Im}(\omega)|$ change significantly. At $\zeta=1$, the variation of the QNFs with $Q$ prior to the outburst slows. As $\zeta$ increases further, the amplitude of $\mathrm{Re}(\omega)$ variation with $Q$ grows, while the behavior of $|\mathrm{Im}(\omega)|$ transitions: for $\zeta=0$ and $1$, it first increases slowly, then decreases, and finally enters an outburst-oscillation stage; whereas for $\zeta=3$, it increases monotonically before entering the outburst stage [see Fig.~\ref{metric1_Re/Im_detail_a}]. For $l=1$, the effect of $\zeta$ on the lower-order overtones remains relatively weak. At $\zeta=1$, only the $n=3$ overtone shows a visible deviation compared to the $\zeta=0$ case [see Fig.~\ref{metric1_Re/Im_detail_d}]. As $\zeta$ increases further, a more pronounced variation of $\mathrm{Re}(\omega)$ with $Q$, similar to that observed for $l=0$, also emerges in the $n=1$ overtones [see Fig.~\ref{metric1_Re/Im_detail_c}]. Furthermore, in general, as $\zeta$ increases, $\mathrm{Re}(\omega)$ decreases for the same $Q$ value, which is opposite to the behavior of the fundamental modes.

\begin{figure}[htbp]
	\centering
	\begin{minipage}{0.45\textwidth}
		\includegraphics[width=\textwidth,keepaspectratio]{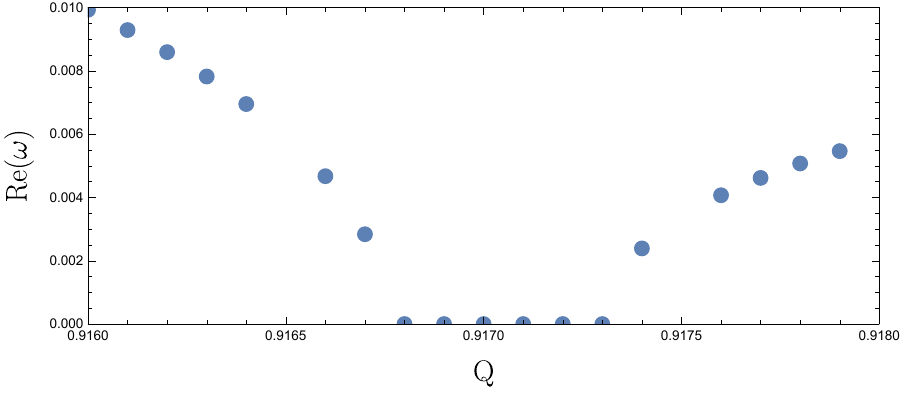}
	\end{minipage}

	\begin{minipage}{0.45\textwidth}
		\includegraphics[width=\textwidth,keepaspectratio]{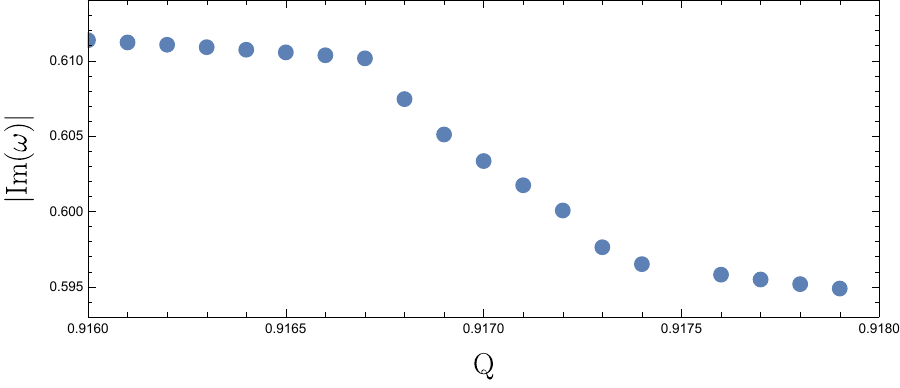}
	\end{minipage}

	\caption{The QNFs of the first overtone modes ($n=1$) for Solution 1 are presented as a function of the charge $Q$ near the purely imaginary region, with parameters set to $l=0$ and $\zeta=3$.}
	\label{metric1_r-i-r}
\end{figure}

It is worth noting that Fig.~\ref{metric1_l=0,1_Re-Im_b} and Fig.~\ref{metric1_l=0,1_Re-Im_c} show that for $l=0$, $n=2$, $\zeta=1$, and for $l=0$, $n=1$, $\zeta=3$, the real part $\mathrm{Re}(\omega)$ for certain overtones decreases rapidly with increasing $Q$, approaching the imaginary axis before moving away again. This behavior reflects a continuous evolution of the QNFs in the complex plane, where the modes undergo a transition from complex modes to purely imaginary modes and then return to the complex modes. Since these frequencies can be continuously tracked throughout the process without any numerical discontinuities near the imaginary axis (see Fig.~\ref{metric1_r-i-r}), this feature is interpreted as a genuine structure of the spectrum rather than a numerical artifact. Similar behaviors have been reported previously in studies of QNMs of the Schwarzschild-dS black hole under gravitational perturbations~\cite{Konoplya:2022xid}.

Moreover, for the second complex overtones with $l=0$ and $\zeta=1$, the modes approach the imaginary axis again at higher $Q$. For $\zeta=3$, due to numerical limitations, the results are presented only up to $Q=0.93$. From the available (but not explicitly shown) calculations, it can be inferred that similar reattachment behaviors may occur in subsequent intervals of $Q$. This recurrent appearance of purely imaginary segments is primarily attributed to the highly oscillatory structure of the overtone spectrum and its modulation by the parameters $\zeta$ and $Q$. In summary, these purely imaginary phenomena are ultimately rooted in the presence of the cosmological constant $\Lambda$, which allows purely imaginary modes to exist in the QNFs and plays a crucial role in shaping the oscillatory structure of the overtones.

\subsubsection{Solution 2}

\begin{figure*}[htbp]
	\centering
	\subfigure[$\zeta=0$ (RN-dS black hole)]{\begin{minipage}{0.33\textwidth}
			\includegraphics[width=\textwidth,keepaspectratio]{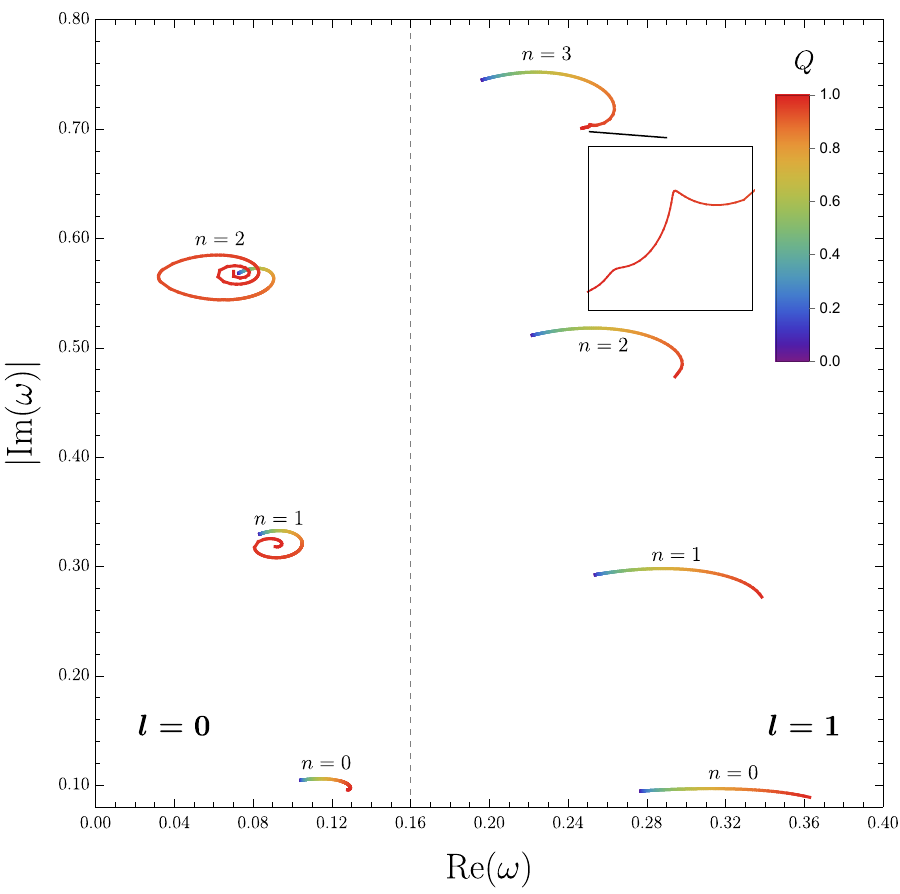}
		\end{minipage}\label{metric2_l=0,1_Re-Im_a}}
	\hspace{-0.55em}
	\subfigure[$\zeta=1$]{\begin{minipage}{0.33\textwidth}
			\includegraphics[width=\textwidth,keepaspectratio]{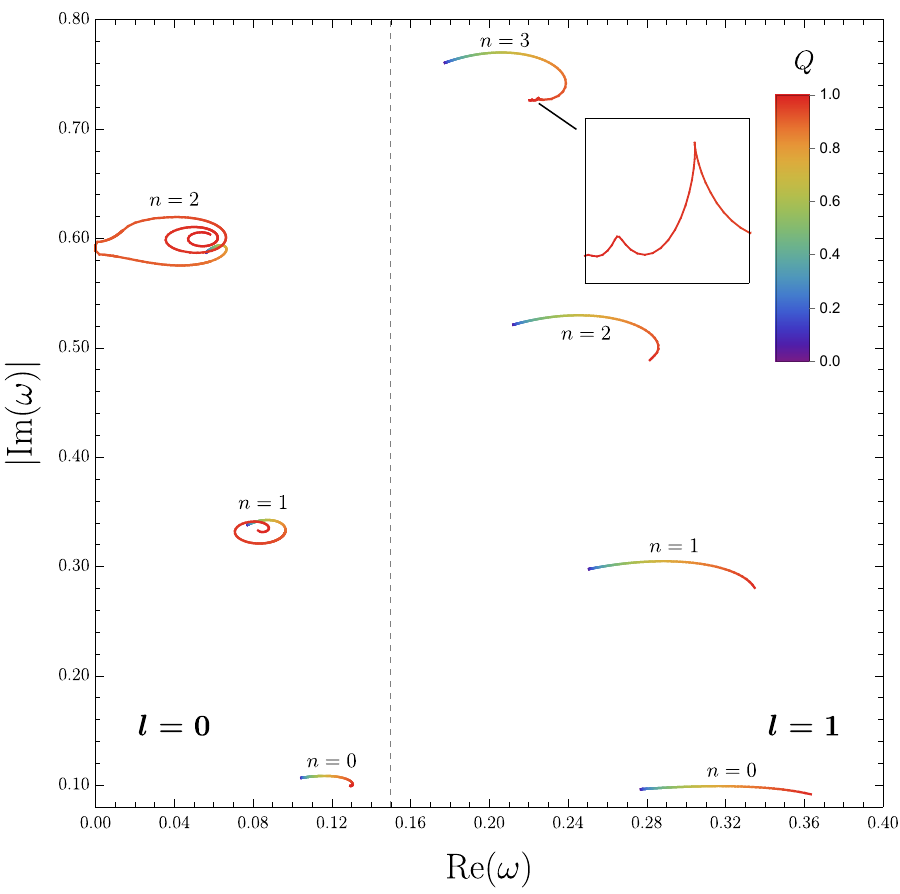}
		\end{minipage}\label{metric2_l=0,1_Re-Im_b}}
	\hspace{-0.55em}
	\subfigure[$\zeta=3$]{\begin{minipage}{0.33\textwidth}
			\includegraphics[width=\textwidth,keepaspectratio]{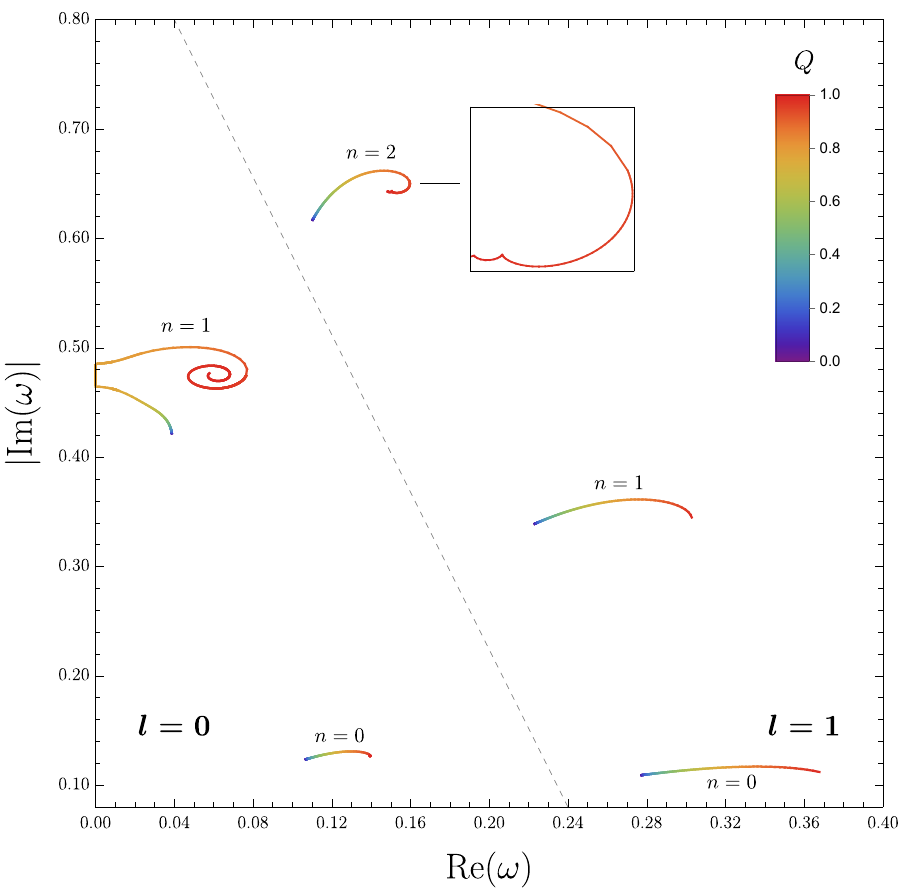}
		\end{minipage}\label{metric2_l=0,1_Re-Im_c}}

	\caption{Phase diagrams $\mathrm{Re}(\omega)$-$|\mathrm{Im}(\omega)|$ of the complex QNFs for Solution 2, which show the fundamental modes and the first few overtone modes. Results are presented for $\Lambda = 0.01$ and $l=0,1$}
	\label{metric2_l=0,1_Re-Im}
\end{figure*}
\begin{figure}[htbp]
	\centering
	\subfigure[$l=0$]{\begin{minipage}{0.22\textwidth}
		\centering
		\includegraphics[width=\textwidth,keepaspectratio]{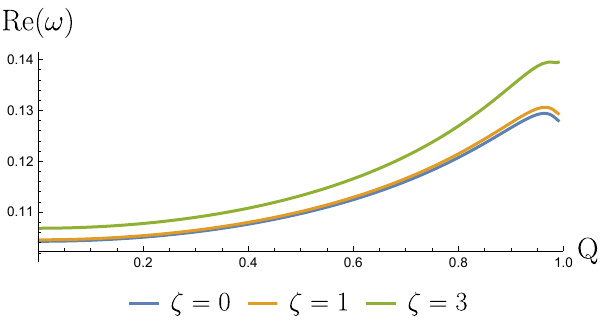}\\[2ex]
		\includegraphics[width=\textwidth,keepaspectratio]{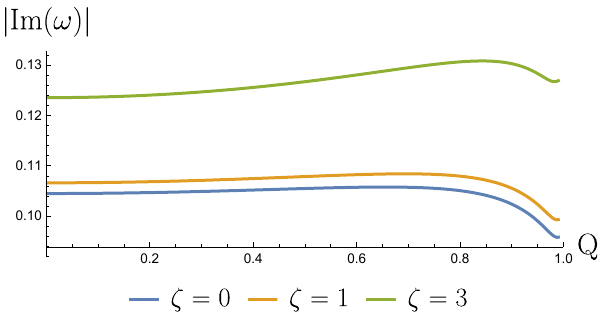}
	\end{minipage}\label{metric2_Re/Im_fundamental_a}}
	\subfigure[$l=1$]{\begin{minipage}{0.22\textwidth}
		\centering
		\includegraphics[width=\textwidth,keepaspectratio]{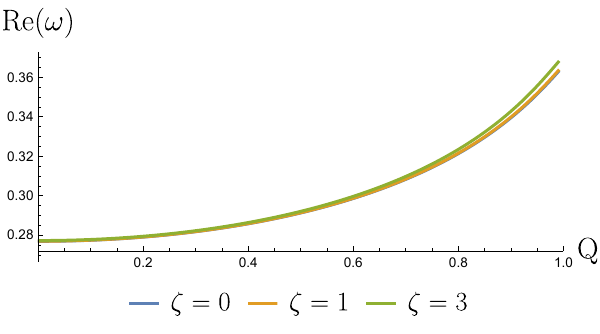}\\[2ex]
		\includegraphics[width=\textwidth,keepaspectratio]{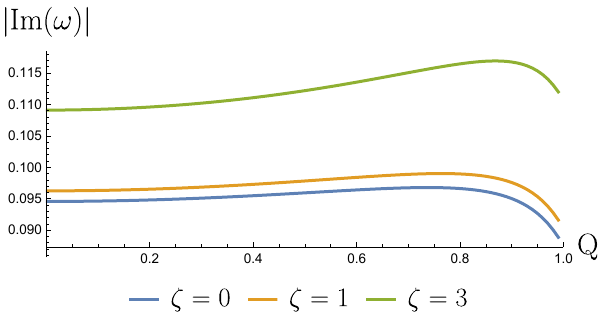}
	\end{minipage}\label{metric2_Re/Im_fundamental_b}}

	\caption{The QNFs of fundamental modes for Solution 2 are presented as a function of the charge $Q$ for $l=0$ and $l=1$.}\label{metric2_Re/Im_fundamental}
\end{figure}
\begin{figure}[htbp]
	\centering
	\subfigure[$l=0,~n=1$]{\begin{minipage}{0.22\textwidth}
		\includegraphics[width=\textwidth,keepaspectratio]{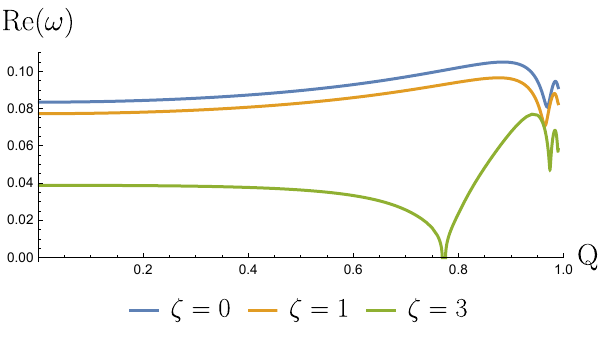}\\[2ex]
		\includegraphics[width=\textwidth,keepaspectratio]{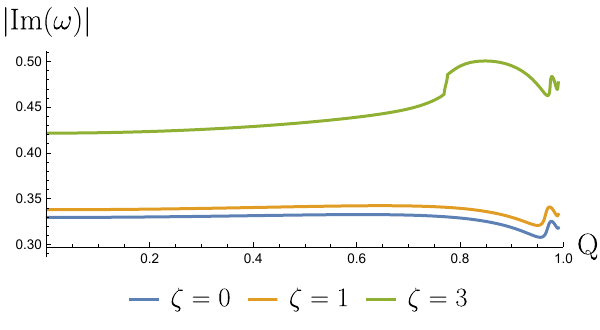}
	\end{minipage}\label{metric2_Re/Im_detail_a}}
	\subfigure[$l=0,~n=2$]{\begin{minipage}{0.22\textwidth}
		\includegraphics[width=\textwidth,keepaspectratio]{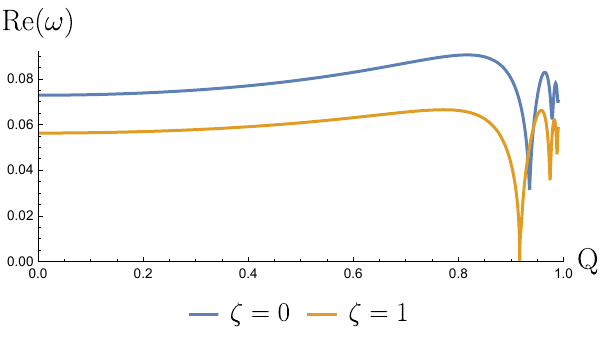}\\[2ex]
		\includegraphics[width=\textwidth,keepaspectratio]{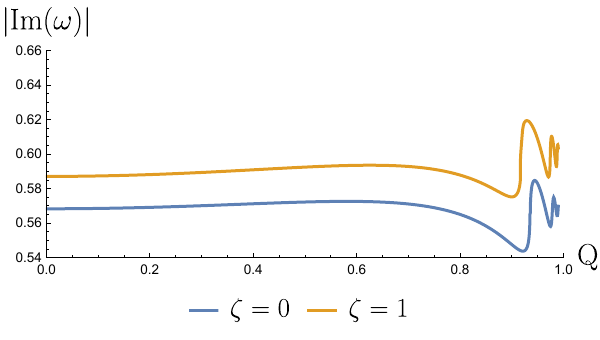}
	\end{minipage}\label{metric2_Re/Im_detail_b}}
	\subfigure[$l=1,~n=2$]{\begin{minipage}{0.22\textwidth}
		\includegraphics[width=\textwidth,keepaspectratio]{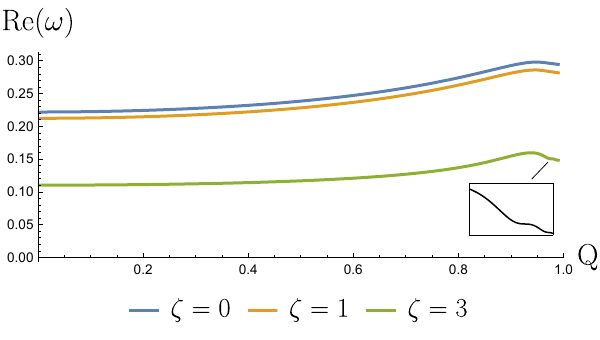}\\[2ex]
		\includegraphics[width=\textwidth,keepaspectratio]{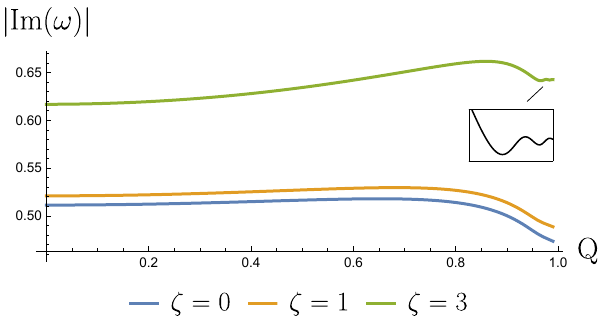}
	\end{minipage}\label{metric2_Re/Im_detail_c}}
	\subfigure[$l=1,~n=3$]{\begin{minipage}{0.22\textwidth}
		\includegraphics[width=\textwidth,keepaspectratio]{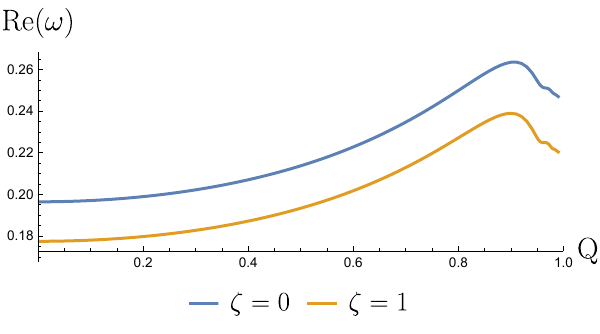}\\[2ex]
		\includegraphics[width=\textwidth,keepaspectratio]{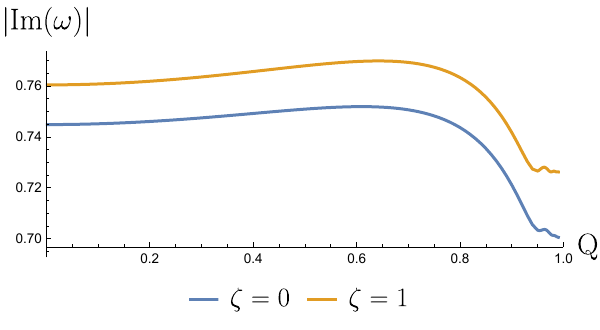}
	\end{minipage}\label{metric2_Re/Im_detail_d}}

	\caption{The QNFs of the first few overtone modes for Solution 2 are presented as a function of the charge $Q$ for $l=0$ and $l=1$.}
	\label{metric2_Re/Im_detail}
\end{figure}
\begin{figure}[htbp]
	\centering
	\begin{minipage}{0.45\textwidth}
		\includegraphics[width=\textwidth,keepaspectratio]{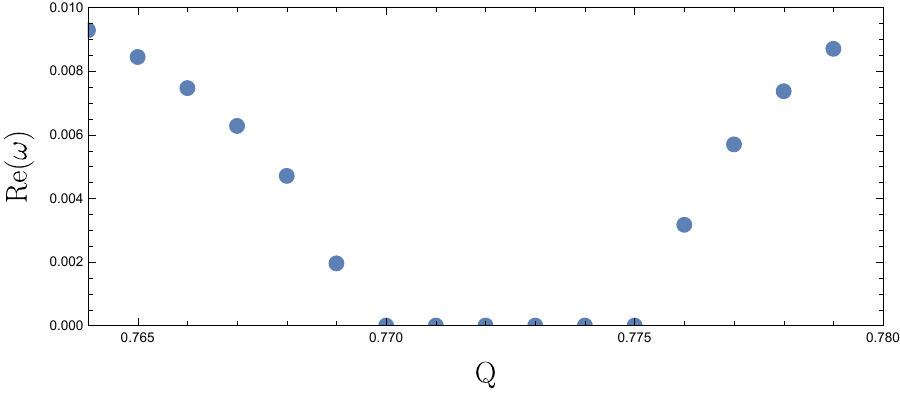}
	\end{minipage}
	\begin{minipage}{0.45\textwidth}
		\includegraphics[width=\textwidth,keepaspectratio]{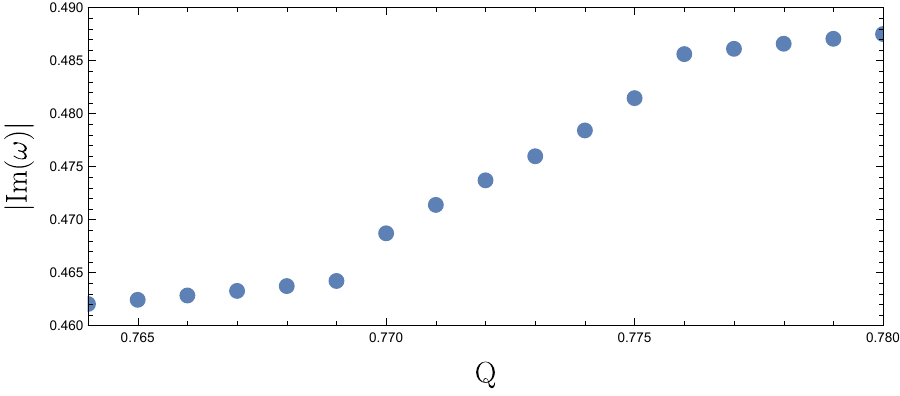}
	\end{minipage}

	\caption{The QNFs of the first complex overtone modes($n=1$) for Solution 2 are presented as a function of the charge $Q$ near the purely imaginary region, with parameters set to $l=0$ and $\zeta=3$.}
	\label{metric2_r-i-r}
\end{figure}
The results of Solution 2, Eq.~\eqref{metric2}, are shown in Fig.~\ref{metric2_l=0,1_Re-Im}. Similarly, we first discuss the fundamental complex modes, as presented in Fig.~\ref{metric2_Re/Im_fundamental}. For $l=0$, similar to Solution 1, the difference in QNFs between $\zeta=0$ and $\zeta=1$ is minor. Both the real part $\mathrm{Re}(\omega)$ and the absolute value of the imaginary part $|\mathrm{Im}(\omega)|$ exhibit non-monotonic behavior with $Q$, initially increasing with $Q$ and then decreasing. As the quantum parameter $\zeta$ further increases to $3$, this non-monotonic behavior remains pronounced, indicating that the increase of $\zeta$ does not eliminate this feature in the present model, in contrast to the suppression observed in Solution 1.

For $l=1$, there are also significant differences between the two solutions. In this model, the effect of the quantum parameter $\zeta$ on $\mathrm{Re}(\omega)$ and $|\mathrm{Im}(\omega)|$ is noticeably smaller than in Solution 1, indicating that the suppression of quantum-gravity effects by the angular quantum number $l$ is stronger in Solution 2 than in Solution 1. Moreover, this stronger suppression does not lead to any significant change in the non-monotonic trend at $\zeta=3$. This further indicates that the influence of $\zeta$ in this model is weak and becomes noticeable only in higher-order overtones. Overall, within Solution 2, the fundamental modes exhibit similar qualitative behavior for different values of $\zeta$, with only quantitative differences.

The detailed results for the complex overtones are shown in Fig.~\ref{metric2_Re/Im_detail}. For $l=0$, the non-monotonic behavior is also enhanced into a pronounced outburst with oscillations. When the quantum parameter $\zeta$ increases to $1$, the trends of the real part $\mathrm{Re}(\omega)$ and the absolute value of the imaginary part $|\mathrm{Im}(\omega)|$ are similar to those in Solution 1, namely, their variation with $Q$ is relatively slow before the outburst stage, while the oscillation intensity after the outburst increases. As $\zeta$ further increases, $\mathrm{Re}(\omega)$ no longer increases monotonically before entering the outburst phase, but instead decreases initially with increasing $Q$. After undergoing a single transition through purely imaginary modes, it increases again and then enters the outburst stage. Correspondingly, $|\mathrm{Im}(\omega)|$ exhibits a rapid increase in the same interval [Fig.~\ref{metric2_Re/Im_detail_a}], and the details of this parameter range are shown in Fig.~\ref{metric2_r-i-r}. Notably, in Solution 1 at $l=0$ and $\zeta=1$, the transition between complex modes and purely imaginary modes occurs twice, whereas under the same parameter conditions this transition occurs only once in Solution 2 [Fig.~\ref{metric2_Re/Im_detail_b}]. This indicates that, at $\zeta = 1$, although the influence of quantum gravity effects on the spectra of the two solutions is difficult to manifest in the fundamental modes, clear differences can still be observed in the overtones. In particular, for Solution 1, the $\mathrm{Re}(\omega)$ of the overtones attains smaller values due to the effect of $\zeta$, making them more susceptible to being captured by the imaginary axis.

For $l=1$, after introducing the quantum parameter $\zeta$, the variations in $\mathrm{Re}(\omega)$ and $|\mathrm{Im}(\omega)|$ for $n \leq 2$ remain minor, showing only small numerical differences. The oscillatory behavior near the extremal regime appears only at $n=3$ for $\zeta=0$ and $1$ [Fig.~\ref{metric2_Re/Im_detail_d}]. Interestingly, when $\zeta$ increases to $3$, these oscillations already appear at $n=2$ [Fig.~\ref{metric2_Re/Im_detail_c}]. This indicates that, although in Solution 2 the enhancement of $\zeta$ on the peculiar spectral behavior is relatively weak and is further suppressed by the angular quantum number $l$, it can still manifest in the QNFs when $\zeta$ reaches $3$.

Overall, the quantum parameter $\zeta$ affects the complex modes differently in the two solutions. Moreover, transitions between complex and purely imaginary modes are observed in both cases. Beyond their dependence on $\zeta$ and $Q$, these transitions are closely related to the intricate interactions between overtones and nearby purely imaginary modes. A clearer understanding of this behavior therefore requires an examination of the complete QNF spectrum.

\subsection{Purely imaginary modes}

In asymptotically dS spacetime, purely imaginary modes are a key feature of the QNMs. They govern the late-time decay of perturbations, producing an exponential profile, and also interact with complex overtones and contribute to the overall spectral structure. In the following analysis, we focus on how the quantum parameter $\zeta$ modifies the decay rates and spectral dominance of purely imaginary modes, as well as their interplay with complex modes. Because purely imaginary modes are also included in the spectrum, the standard QNM index $n$ is no longer sufficient to characterize all modes. We therefore introduce $\tilde{n}$, which labels the modes in ascending order of $|\mathrm{Im}(\omega)|$ and thus reflects their relative damping strengths. Unlike the conventional overtone index $n$, the index $\tilde{n}$ does not correspond to a specific mode branch. Tables~\ref{pure1} and \ref{pure2} present the complete QNFs, including both complex and purely imaginary modes, for $l=0$ and $\Lambda=0.01$ at different values of $\zeta$ and $Q$ for the two solutions. The tabulated results show that increasing the quantum parameter $\zeta$ gradually changes the relative dominance between complex and purely imaginary modes. In both solutions, when $\zeta$ is small, the lowest-lying QNMs are complex, indicating that the early-time evolution of the perturbation is dominated by oscillatory decay for $\Lambda=0.01$. As $\zeta$ increases, purely imaginary modes become increasingly dominant. For example, in Solution 1 with $\zeta=3$ and sufficiently large $Q$, the first two lowest-lying modes become purely imaginary. In this regime, the oscillatory stage terminates earlier, and the signal will transition more rapidly into the non-oscillatory exponential decay stage.

Moreover, comparison between the two solutions shows that the dominance of purely imaginary modes grows more slowly with increasing $\zeta$ in Solution 2 than in Solution 1. At the same time, as purely imaginary modes become increasingly dominant, fewer complex modes are available within the spectral range computable by the PSM. As a result, extracting higher-order complex overtones becomes progressively more difficult. Overall, purely imaginary modes form an essential part of the spectrum, significantly influencing the early-time decay of perturbations and playing a key role in the computation of higher-order complex overtones.

\begin{figure}[htbp]
	\centering
	\begin{minipage}{0.45\textwidth}
		\includegraphics[width=\textwidth,keepaspectratio]{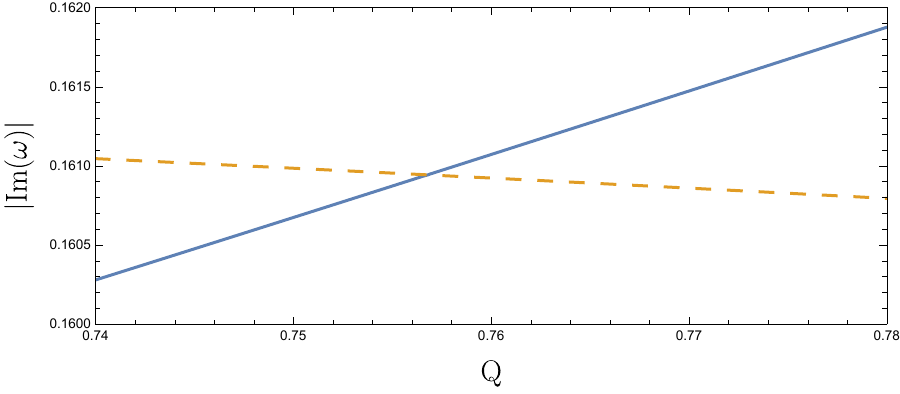}
	\end{minipage}

	\caption{The damping-rate crossing of QNFs for Solution 1 as a function of the charge $Q$, with parameters set to $l=0$ and $\zeta=3$, including the fundamental complex modes (blue line) and the nearby purely imaginary modes (orange dashed line)}
	\label{case3}
\end{figure}

\begin{figure}[htbp]
	\centering
	\begin{minipage}{0.45\textwidth}
		\includegraphics[width=\textwidth,keepaspectratio]{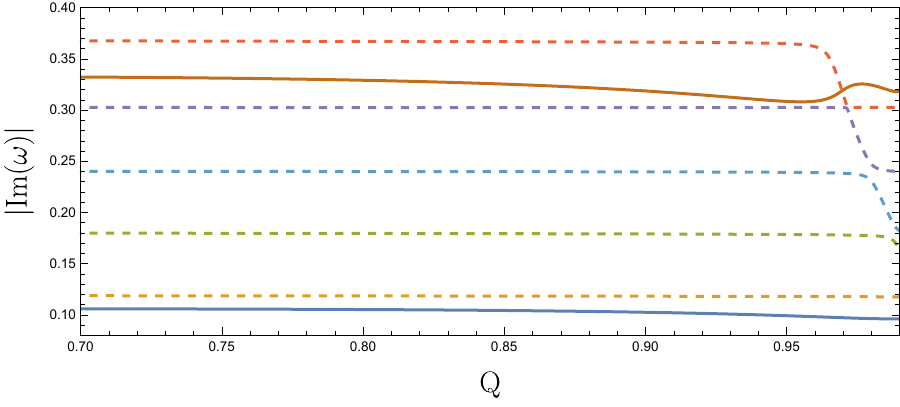}
	\end{minipage}

	\caption{The damping-rate crossing of QNFs for Solution 1 as a function of the charge $Q$, with parameters set to $l=0$ and $\zeta=0$. The blue and brown lines respectively represent the fundamental complex modes and the first complex overtone modes, while the other dashed lines represent the nearby purely imaginary modes.}
	\label{case2}
\end{figure}

\begin{table*}[htbp]
	\centering
	\fontsize{9}{9}\selectfont
	\begin{tabular}{|c|c|c|c|c|}
		\hline
		$Q$ & $\tilde{n}$ & $\zeta=1$ & $\zeta=2$ & $\zeta=3$ \\
		\hline

		\multirow{4}{*}{0}
		& 0 & 0.10484$-$0.10880\,i & $-$0.11525\,i & $-$0.10997\,i \\
		& 1 & $-$0.11821\,i & 0.10632$-$0.12224\,i & 0.10871$-$0.14635\,i \\
		& 2 & $-$0.17860\,i & $-$0.17300\,i & $-$0.16300\,i \\
		& 3 &$-$0.23678\,i &$-$0.22678\,i &$-$0.21371\,i \\
		\hline

		\multirow{4}{*}{0.3}
		& 0 & 0.10676$-$0.10938\,i & $-$0.11514\,i & $-$0.10985\,i \\
		& 1 & $-$0.11810\,i & 0.10843$-$0.12337\,i & 0.11113$-$0.14847\,i \\
		& 2 & $-$0.17848\,i & $-$0.17282\,i & $-$0.16271\,i \\
		& 3 &$-$0.23673\,i &$-$0.22664\,i &$-$0.21339\,i \\
		\hline

		\multirow{4}{*}{0.6}
		& 0 & 0.11339$-$0.11081\,i & $-$0.11479\,i & $-$0.10950\,i \\
		& 1 & $-$0.11777\,i & 0.11587$-$0.12672\,i & 0.1977$-$0.15528\,i \\
		& 2 & $-$0.17805\,i & $-$0.17219\,i & $-$0.16177\,i \\
		& 3 &$-$0.23652\,i &$-$0.22608\,i &$-$0.21226\,i \\
		\hline

		\multirow{4}{*}{0.9}
		& 0 & 0.12854$-$0.10880\,i & $-$0.11410\,i & $-$0.10886\,i \\
		& 1 & $-$0.11711\,i & 0.13372$-$0.12910\,i & $-$0.15985\,i \\
		& 2 & $-$0.17685\,i & $-$0.17051\,i & 0.14116$-$0.16643\,i \\
		& 3 &$-$0.23558\,i &$-$0.22370\,i &$-$0.20933\,i \\
		\hline
	\end{tabular}
	\centering

	\caption{Complete QNFs (including both complex and purely imaginary modes) of Solution 1 for $l=0$ and $\Lambda = 0.01$, obtained via PSM, are presented for various values of the charge $Q$ and quantum parameter $\zeta$.}\label{pure1}
\end{table*}
\begin{table*}[htbp]
	\centering
	\fontsize{9}{9}\selectfont
	\begin{tabular}{|c|c|c|c|c|}
		\hline
		$Q$ & $\tilde{n}$ & $\zeta=1$ & $\zeta=2$ & $\zeta=3$ \\
		\hline

		\multirow{4}{*}{0}
		& 0 & 0.10457$-$0.10662\,i & 0.10541$-$0.11298\,i &$-$0.11467\,i \\
		& 1 &$-$0.11868\,i &$-$0.11723\,i & 0.10684$-$0.12354\,i \\
		& 2 &$-$0.17949\,i &$-$0.17673\,i &$-$0.17189\,i \\
		& 3 &$-$0.23853\,i &$-$0.23328\,i &$-$0.22534\,i \\
		\hline

		\multirow{4}{*}{0.3}
		& 0 & 0.106449$-$0.10712\,i & 0.10740$-$0.11372\,i &$-$0.11456\,i \\
		& 1 &$-$0.11857\,i &$-$0.11712\,i & 0.10900$-$0.12468\,i \\
		& 2 &$-$0.17938\,i &$-$0.17660\,i &$-$0.17170\,i \\
		& 3 &$-$0.23848\,i &$-$0.23321\,i &$-$0.22518\,i \\
		\hline

		\multirow{4}{*}{0.6}
		& 0 & 0.11294$-$0.10824\,i & 0.11431$-$0.11571\,i & $-$0.11420\,i \\
		& 1 & $-$0.11825\,i & $-$0.11678\,i & 0.11658$-$0.12805\,i \\
		& 2 & $-$0.17898\,i & $-$0.17610\,i & $-$0.17103\,i \\
		& 3 &$-$0.23831\,i &$-$0.23289\,i &$-$0.22454\,i \\
		\hline

		\multirow{4}{*}{0.9}
		& 0 & 0.12759$-$0.10555\,i & 0.13037$-$0.11492\,i & $-$0.11352\,i \\
		& 1 & $-$0.11759\,i & $-$0.11611\,i & 0.13476$-$0.13029\,i \\
		& 2 & $-$0.17786\,i & $-$0.17474\,i & $-$0.16931\,i \\
		& 3 &$-$0.23758\,i &$-$0.23148\,i &$-$0.22209\,i \\
		\hline
	\end{tabular}
	\centering

	\caption{Complete QNFs (including both complex and purely imaginary modes) of Solution 2 for $l=0$ and $\Lambda = 0.01$, obtained via PSM, are presented for various values of the charge $Q$ and quantum parameter $\zeta$.}\label{pure2}
\end{table*}
It is worth noting that special interactions arise between complex modes and nearby purely imaginary modes. From the tables, when $Q$ is small, the $|\mathrm{Im}(\omega)|$ of different modes vary monotonically with $Q$. However, the fundamental complex mode and its nearby purely imaginary mode exhibit opposite trends, causing their $|\mathrm{Im}(\omega)|$ to approach each other and, in some parameter ranges, even cross (see Fig.~\ref{case3}). In the RN-dS black hole, purely imaginary modes always exist, but within the parameter range we consider, their dominance is relatively weak, making such damping-rate crossings less pronounced. Nevertheless, as shown in Fig.~\ref{case2}, even for $\zeta=0$, certain purely imaginary modes become highly sensitive to parameter variations during overtone outbursts and undergo rapid changes near complex overtones. Damping-rate crossings can also occur in this regime. This indicates that local damping-rate crossings are not phenomena unique to quantum corrections, but already exist in the overtone structure of RN-dS black holes in a subtler form. The introduction of the quantum parameter $\zeta$ does not alter this underlying mechanism of these crossings. Instead, by enhancing the relative dominance of purely imaginary modes within the spectrum, it allows the damping-rate crossings that originally existed only in the overtones to manifest among the fundamental complex modes, thereby amplifying their physical observability.
Besides damping-rate crossings, complex and purely imaginary modes can also undergo more intricate splitting-merging interactions. The introduction of the quantum parameter $\zeta$ reduces the real part $\mathrm{Re}(\omega)$ of the complex overtones. Combined with the strong spectral variations associated with overtone outbursts, this makes the overtones more likely to approach the imaginary axis and undergo the transitions shown in Figs.~\ref{metric1_r-i-r} and \ref{metric2_r-i-r}.

\begin{figure}[htbp]
	\centering
	\begin{minipage}{0.45\textwidth}
		\includegraphics[width=\textwidth,keepaspectratio]{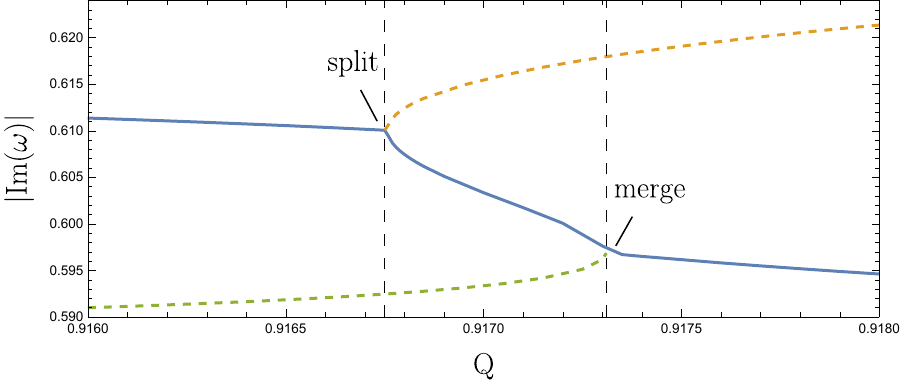}
	\end{minipage}

	\caption{QNFs of the first complex overtone modes (blue line) for Solution 1 as a function of the charge $Q$ near the purely imaginary region, illustrating the splitting and subsequent merging behavior of modes close to the imaginary axis, with parameters set to $l=0$ and $\zeta=3$. The blue branch splits into a new purely imaginary mode branch (orange dashed line) when entering the transition region, and upon leaving the transition region, it merges with an existing purely imaginary mode branch (green dashed line).}
	\label{case4_1}
\end{figure}

\begin{figure}[htbp]
	\centering

	\begin{minipage}{0.45\textwidth}
		\includegraphics[width=\textwidth,keepaspectratio]{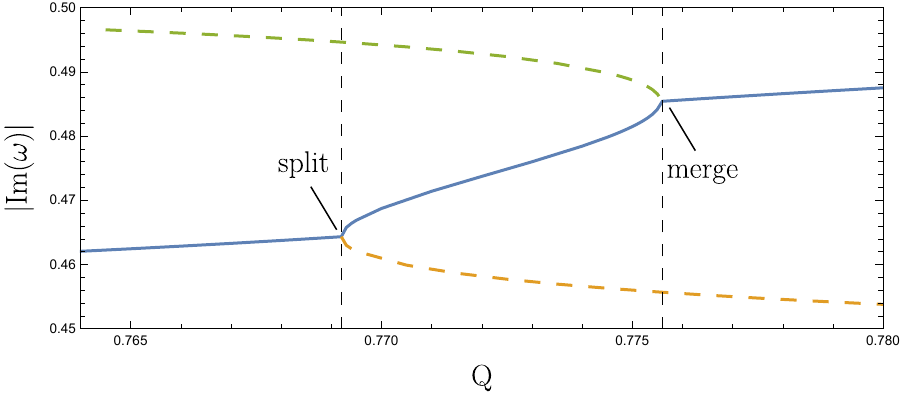}
	\end{minipage}

	\caption{QNFs of the first complex overtone modes (blue lines) for Solution 2 as a function of the charge $Q$ near the purely imaginary region, illustrating the splitting and subsequent merging behavior of modes close to the imaginary axis, with parameters set to $l=0$ and $\zeta=3$. The blue branch splits into a new purely imaginary mode branch (orange dashed line) when entering the transition region, and upon leaving the transition region, it merges with an existing purely imaginary mode branch (green dashed line).}
	\label{case4_2}
\end{figure}

To illustrate these interactions in greater detail, Figs.~\ref{case4_1} and \ref{case4_2} present spectra that explicitly include purely imaginary modes. Since the transition patterns observed in different cases are qualitatively similar, we illustrate them here using Solution 1 as an example, see Fig.~\ref{case4_1}. In this figure, the blue solid line represents complex modes and exhibits three distinct branches within the transition region. The orange dashed line denotes purely imaginary modes that split from the complex branch upon entering the transition region ($Q\approx0.9168$), while the green dashed line corresponds to purely imaginary modes that persist throughout the evolution. Upon leaving the transition ($Q\approx0.9173$), the green branch merges into the blue branch and recovers a nonzero $\mathrm{Re}(\omega)$. This transient purely imaginary behavior reflects nontrivial interactions between complex and purely imaginary modes during overtone outbursts, leading to local rearrangements of the QNF branch structure and further illustrates the intricate organization of the full spectrum.


In summary, the results of this section demonstrate that complex and purely imaginary modes interact nontrivially within the complete QNF spectrum. The presence of the cosmological constant $\Lambda$ permits purely imaginary modes, whose interactions with complex modes lead to a significantly richer spectral structure and highlight the importance of considering the full spectrum rather than isolated mode branches. It is also worth noting that near-extremal parameter regimes provide a particularly sensitive regime for the spectrum, where overtone outbursts, damping-rate crossings, and branch rearrangements occur frequently. This suggests that local spectral rearrangements are closely connected to the dynamics of overtone outbursts. In addition, as the quantum parameter $\zeta$ increases, mode interactions become more frequent and pronounced, making the continuous tracking and identification of individual modes along parameter evolution progressively more difficult. Consequently, when the complete spectrum is taken into account, the conventional overtone-based classification becomes less sharply defined.

\begin{figure}[htbp]
	\centering
	\begin{minipage}{0.45\textwidth}
		\includegraphics[width=\textwidth,keepaspectratio]{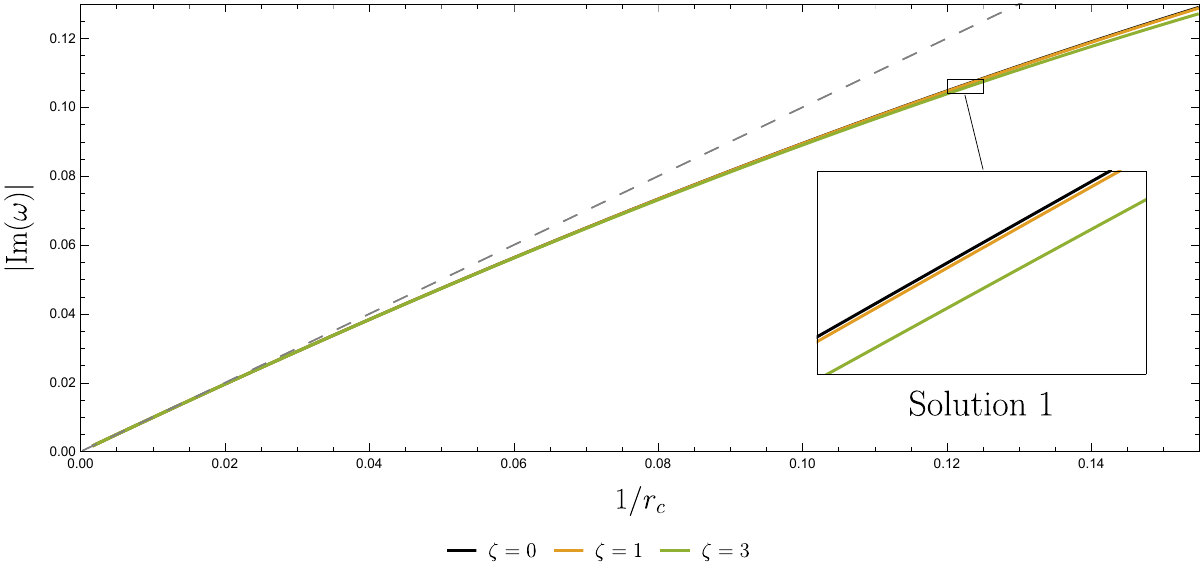}
	\end{minipage}
	\begin{minipage}{0.45\textwidth}
		\includegraphics[width=\textwidth,keepaspectratio]{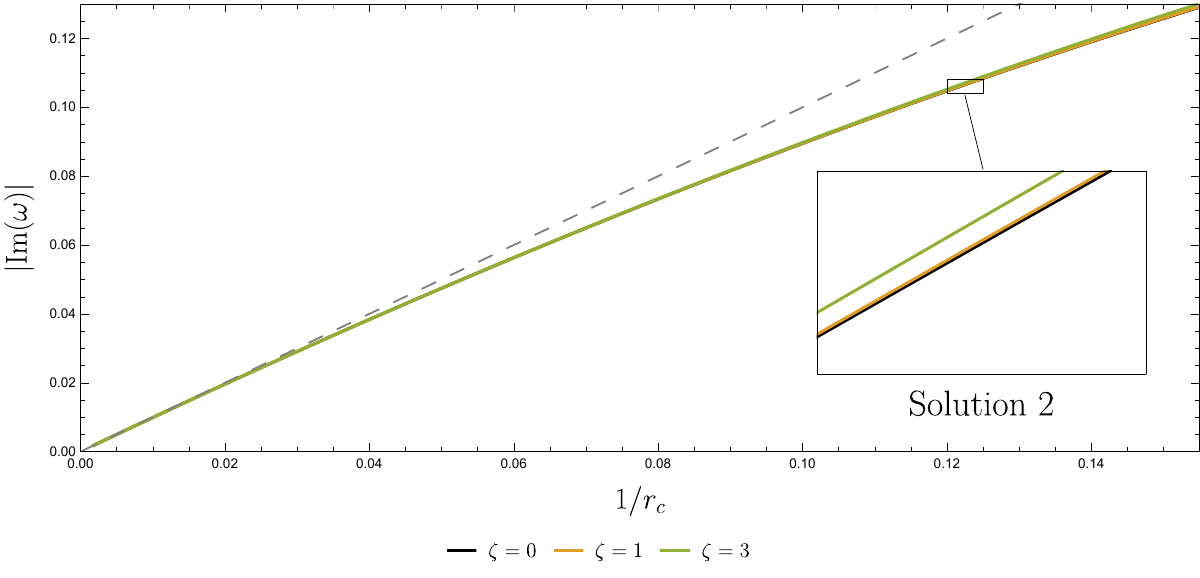}
	\end{minipage}

	\caption{The $|\mathrm{Im}(\omega)|$ of the lowest-lying QNMs as functions of $1/r_c$ for the two solutions, with $l=0$ and $Q=0.5$. The cosmological horizon radius $r_c$ is varied by changing the cosmological constant $\Lambda$. The dashed lines represent the reference scaling $|\mathrm{Im}(\omega)|\propto 1/r_c$.
 }
	\label{rc_i}
\end{figure}

Beyond the spectral interactions discussed above, the purely imaginary modes also provide an important connection to the SCCC. According to the general analysis for arbitrary spherically symmetric asymptotically dS black holes, the purely imaginary dS branch exhibits a universal inverse scaling of the damping rate with the cosmological horizon radius $r_c$ in the regime of $r_c\gg r_h$, which leads to the satisfaction of the SCCC bound for sufficiently small black holes~\cite{Konoplya:2022kld}. In the present models, we vary the cosmological constant $\Lambda$ while keeping the charge $Q$ and quantum parameter $\zeta$ fixed, thereby changing the cosmological horizon radius $r_c$. The $|\mathrm{Im}(\omega)|$ of the lowest-lying QNMs is shown as a function of $1/r_c$ in Fig.~\ref{rc_i}. The results show that, in the regime of large $r_c$ ($1/r_c\rightarrow0$), the damping rates of the dS branch approach the inverse scaling behavior $|{\rm Im}(\omega)|\propto 1/r_c$, consistent with the behavior reported in Ref.~\cite{Konoplya:2022kld}. Moreover, although the quantum parameter $\zeta$ modifies the damping rates quantitatively, they do not alter this behavior.

\section{Conclusion}
\label{section5}

In this work, we systematically investigated the QNFs of two electrically charged black hole solutions with a cosmological constant under massless scalar field perturbations. By analyzing the effective potential, the QNFs, and the decay behavior of perturbations, we explored the influence of the quantum parameter $\zeta$ on the spectral properties of the two solutions. The main results are summarized as follows:
\begin{enumerate}
\item Analysis of the effective potential shows that Solution 1 is more sensitive to the quantum parameter $\zeta$, exhibiting a higher potential barrier, whereas Solution 2 shows a weaker response. In addition, Solution 1 has a larger negative region, suggesting increased susceptibility to instability under extreme conditions.

\item Although increasing the cosmological constant $\Lambda$ weakens the influence of the quantum parameter $\zeta$ on the QNFs, the characteristic spectral features introduced by $\zeta$ are still preserved.

\item For the conventional complex modes, the two solutions exhibit significant differences only when $\zeta$ is sufficiently large. For the fundamental modes, when $\zeta$ reaches 3, the non-monotonic behavior of the QNF spectrum in Solution 1 is suppressed, whereas such suppression is not observed in Solution 2. For the overtones, the quantum gravity effect cannot be simply described as either an enhancement or suppression of overtone outbursts, but instead exhibits additional features. These results indicate a nontrivial interplay between quantum gravity effects and charge. The resulting differences provide unique spectral signatures to distinguish quantum-corrected black holes observationally and to constrain or exclude certain quantization schemes.

\item For purely imaginary modes, increasing $\zeta$ strengthens their dominance and accelerates the transition to exponential decay. Furthermore, complex and purely imaginary modes exhibit interactions that generally accompany overtone outbursts in near-extremal regimes, including damping-rate crossings and splitting--merging behavior, which become more frequent as $\zeta$ increases. Although these effects may complicate the tracking of individual modes, they highlight the importance of analyzing the complete QNF spectrum in revealing overtone outbursts and the deeper spectral consequences of quantum corrections.
\end{enumerate}

To investigate the effects of the black hole charge and quantum parameter on the overtone structure, we consider the simplest case of a massless neutral scalar field. However, other types of perturbations may exhibit qualitatively different features. In particular, for charged black hole spacetimes, charged scalar perturbations deserve special attention, since the interaction between the scalar field charge and the black hole charge may induce superradiant instabilities. In the RN-dS spacetime, the interaction between the charge of the scalar field and that of the black hole can trigger a superradiant instability, which occurs only for the $l=0$ modes, while higher angular quantum numbers $l$ raise the potential barrier and render the field more stable~\cite{Zhu:2014sya,Konoplya:2014lha}. Since quantum corrections can significantly modify the shape of the effective potential, they are expected to further adjust the critical values of this instability, or even suppress the growing modes. Therefore, a systematic study of charged scalar perturbations in the covariant effective quantum black hole backgrounds considered here would provide a new avenue for exploring quantum gravity effects and constitutes a valuable direction for future work.

\begin{acknowledgements}
We would like to thank Lei You for helpful discussions. This work is supported in part by NSFC Grants No. 12165005 and No. 11961131013.
\end{acknowledgements}

\appendix

\section{Pseudo-spectral method}
\label{app:A}

In this appendix, we present a detailed description of the PSM used in Sect.~\ref{section4}. For a more comprehensive introduction, readers may consult~\cite{Jansen:2017oag,Boyd2000} and related literature~\cite{Fu:2022cul,Fu:2023drp,Gong:2023ghh,Zhang:2024nny,Zhu:2024wic}

In Eq.~\eqref{ef_metric}, we presented the metric after the coordinate transformation:
\begin{equation}\label{appef_metric}
	\mathrm{d}s_{EF}^2=-f(u)\mathrm{d}v^2-\frac{2\mathrm{d}v\mathrm{d}u}{u^2\sqrt{g(u)}}+\frac{1}{u^2}\mathrm{d}\Omega^2\,,
\end{equation}
for the asymptotically dS case, the radial coordinate $u$ is defined in the interval $(u_c,u_+)$, where $u_c = 1/r_c$ and $u_+ = 1/r_+$ denote the locations of the cosmological and event horizons in the $u$ coordinate, respectively. To map this interval to the standard domain $(0,1)$ commonly used in the PSM, we introduce:
\begin{equation}
	z=\frac{u-u_c}{u_+-u_c}\,,
\end{equation}
so that the cosmological and event horizons correspond to $z=0$ and $z=1$, respectively. At $z \to 0$ (i.e., $u\to u_c$), a purely outgoing wave boundary condition must be imposed. To achieve this, we implement the following transformation:
\begin{equation}
	\psi(u)=\frac{1}{u-u_c}(u-u_c)^{-i\omega/\kappa_c}\delta\psi(u)\,,
\end{equation}
where the $\kappa_c$ is the surface gravity at the cosmological horizon, defined as
\begin{equation}
		\kappa_c=\left.\left(\frac{u^2}{2}\sqrt{\frac{\mathrm{d} f(u)}{\mathrm{d} u}\frac{\mathrm{d} [f(u)g(u)]}{\mathrm{d} u}}\right)\right|_{u=u_c}\,.
\end{equation}
Combined with the coordinate transformation $z$, the perturbation equation becomes:
\begin{equation}\label{last_function}
	A_0\delta\psi(z)+A_1\delta\psi^\prime(z)+A_2\delta\psi^{\prime\prime}(z)=0\,,
\end{equation}
where $A_0$, $A_1$, and $A_2$ are polynomials of the parameters $Q, z, u_c, u_+, \zeta, \omega$ and $l$ (their explicit forms are omitted due to length). The rewritten equation satisfies the required ingoing and outgoing boundary conditions at $z=1$ and $z=0$, respectively. Here, $Q$, $l$, and $\zeta$ are given parameters, while $u_c$ and $u_+$ are determined numerically from given values of $M$ and $\Lambda$. Consequently, the only unknowns in Eq.~\eqref{last_function} are the radial coordinate $z$ and the QNFs $\omega$.

The key to the PSM is to discretize the linear differential equation Eq.~\eqref{last_function}, and then solve the resulting generalized eigenvalue problem. The first step in this process is to replace the continuous variables with a discrete set of collocation points, known as the grid points. Through this step, a function can be represented by its values at these grid points; in particular, $\delta\psi(z)$ in Eq.~\eqref{last_function} can be expressed as:
\begin{equation}
	\delta\psi(z)=\sum_{j=0}^{N}\alpha(z_j)C_j(z)\,,
\end{equation}
where $z_j$ are the grid points, and the polynomial $C_j(z)$ is the basis function associated with the grid points, satisfying $C_j(z_i)=\delta_{ij}$ for $i,j=0,\ldots,N$. The choice of a specific grid uniquely determines the cardinal functions $C_j(z)$. They are given by:
\begin{equation}
	C_j(z)=\prod_{j=0,j\neq i}^{N}\frac{z-z_j}{z_i-z_j}\,,
\end{equation}
Next, a matrix $D^{(1)}_{ij}$ representing the first derivative is constructed using the basis functions, defined by $D^{(1)}_{ij}=C_i^\prime(z_j)$. Similarly, matrices for higher-order derivatives can be further constructed.

The successful implementation of this method critically depends on the choice of collocation points. These points determine the accuracy and convergence of the spectral approximation. Among various options, the Chebyshev grid is generally preferred and frequently used due to its superior performance. The Chebyshev grid is defined as:
\begin{equation}
	z_i=\cos\Big(\frac{i}{N}\pi\Big)\,.
\end{equation}
For this set of grid points, the basis functions $C_j(z)$ can be expressed as a linear combination of Chebyshev polynomials $T_n(x)$:
\begin{equation}
	C_j(z)=\frac{2}{Np_j}\sum_{m=0}^{N}\frac{1}{p_m}T_m(z_j)T_m(z)\,,
\end{equation}
where $p_j$ are weighting factors. Specifically, $p_j=2$ when $j=0$ or $j=N$, and $p_j=1$ when $j\neq0$ and $j\neq N$. These weighting factors play a key role in ensuring numerical stability. The basis functions $C_j(z)$ occupy a central role in the discretization process, as they provide smooth and accurate basis functions for function approximation on the Chebyshev grid.

In general, the coefficients $A_k$ ($k=0,1,2$) are polynomials of the frequency $\omega$, which can be expanded as $A_k=\sum_P c_{k,P}(z)\omega^P$. For computational convenience, the coefficients $c_{k,P}(z)$ are discretized at the grid points, thereby converted into vectors. These vectors are then multiplied by the corresponding derivative matrices $D^{(n)}_{ij}$, and the resulting matrices are summed. This procedure allows equation Eq.~\eqref{last_function} to be reformulated as the following generalized eigenvalue equation:
\begin{equation}
	\begin{aligned}\label{matrix_eq2}
		\mathcal{M}(\omega)\delta\psi&=(M_0+\omega M_1+\omega^2M_2+\cdots+\omega^PM_P)\delta\psi\\
		&=0\,.
	\end{aligned}
\end{equation}
This is exactly the form of Eq.~\eqref{matrix_eq} in the main text, where $M_P$ is a linear combination of derivative matrices and represents a purely numerical matrix, with $(M_P)_{ij}=c_{0,P}(z_i)\delta_{ij}+c_{1,P}(z_i)D^{(1)}_{ij}+c_{2,P}(z_i)D^{(2)}_{ij}$. Through the above transformations, the linear ordinary differential equation \eqref{last_function} is converted into a matrix equation under specific boundary conditions. By numerically solving the eigenvalue equation \eqref{matrix_eq2}, the QNFs can be determined under the specified boundary conditions.

\section{The results with larger $\Lambda$}
\label{app:B}

\begin{figure}[htbp]
	\centering
	\subfigure[$\Lambda=0.05$]{\begin{minipage}{0.22\textwidth}
			\centering
			\includegraphics[width=\textwidth,keepaspectratio]{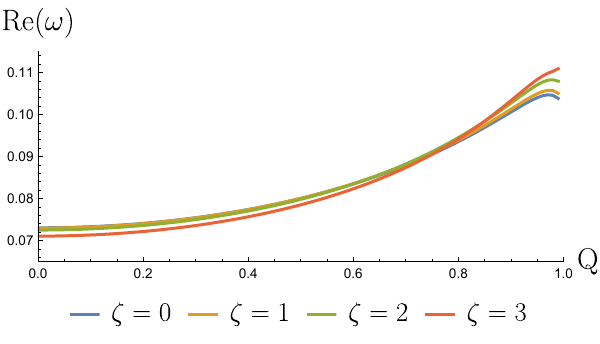}\\[2ex]
			\includegraphics[width=\textwidth,keepaspectratio]{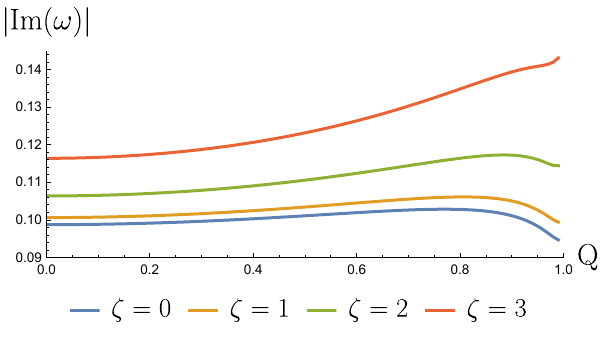}
		\end{minipage}\label{metric1_l=0_variate:Q_a}}
	\subfigure[$\Lambda=0.1$]{\begin{minipage}{0.22\textwidth}
			\centering
			\includegraphics[width=\textwidth,keepaspectratio]{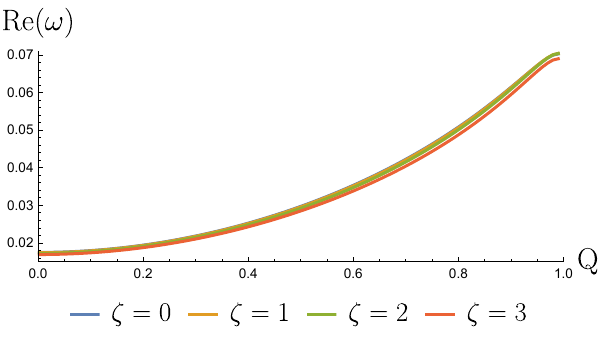}\\[2ex]
			\includegraphics[width=\textwidth,keepaspectratio]{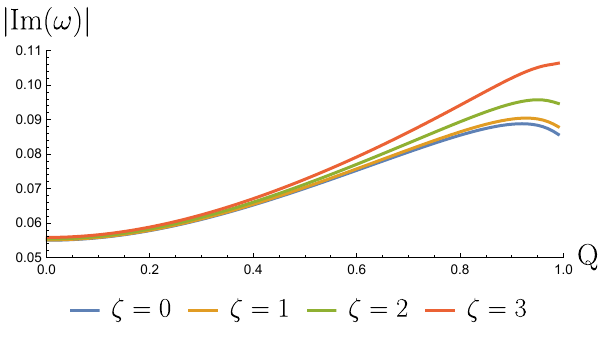}
		\end{minipage}\label{metric1_l=0_variate:Q_b}}

	\caption{Fundamental complex QNFs for Solution 1 are presented as a function of the charge $Q$ for $\Lambda=0.05$ and $\Lambda=0.1$.}
	\label{metric1_l=0_variate:Q}
\end{figure}

\begin{figure}[htbp]
	\centering
	\subfigure[$\Lambda=0.05$]{\begin{minipage}{0.22\textwidth}
			\centering
			\includegraphics[width=\textwidth,keepaspectratio]{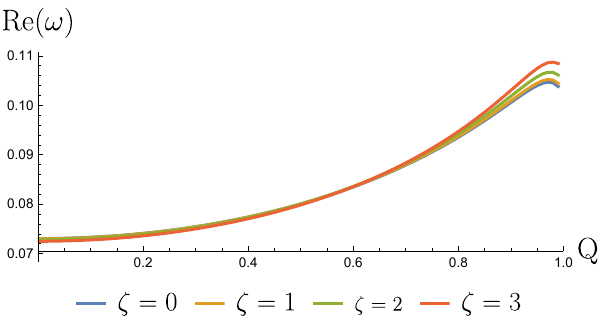}\\[2ex]
			\includegraphics[width=\textwidth,keepaspectratio]{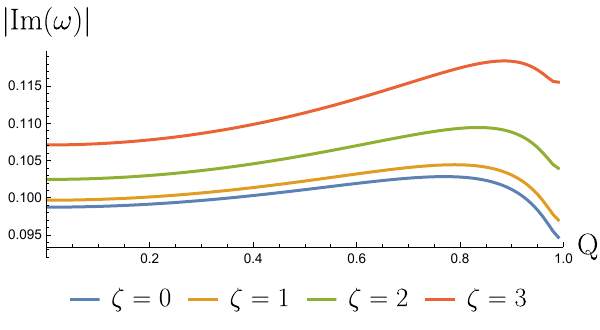}
		\end{minipage}\label{metric2_l=0_variate:Q_a}}
	\subfigure[$\Lambda=0.1$]{\begin{minipage}{0.22\textwidth}
			\centering
			\includegraphics[width=\textwidth,keepaspectratio]{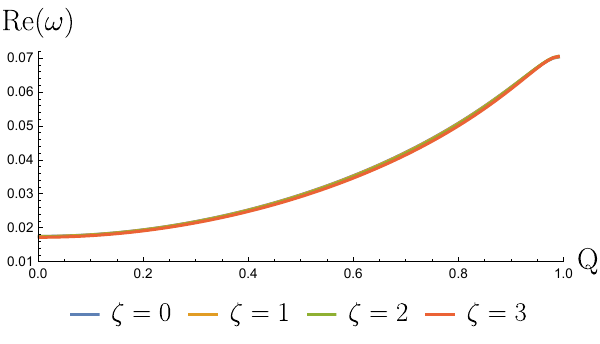}\\[2ex]
			\includegraphics[width=\textwidth,keepaspectratio]{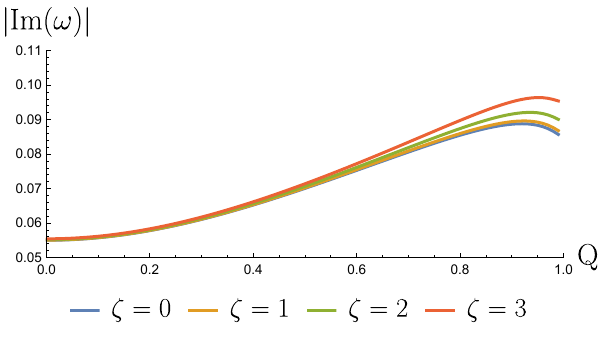}
		\end{minipage}\label{metric2_l=0_variate:Q_b}}

	\caption{Fundamental complex QNFs for Solution 2 are presented as a function of the charge $Q$ for $\Lambda=0.05$ and $\Lambda=0.1$.}
	\label{metric2_l=0_variate:Q}
\end{figure}

As $\Lambda$ increases, both the real parts $\mathrm{Re}(\omega)$ and the absolute values of the imaginary parts $|\mathrm{Im}(\omega)|$ of the QNFs undergo significant changes, indicating the direct influence of the cosmological constant on the spectrum. As shown in Figs.~\ref{metric1_l=0_variate:Q} and \ref{metric2_l=0_variate:Q}, for larger $\Lambda$ and small $Q$, the spectral separation corresponding to different values of the quantum parameter $\zeta$ becomes less pronounced, with the corresponding curves lying closer together. This indicates that, at a numerical level, increasing $\Lambda$ reduces the separation between different $\zeta$ sectors.

However, as $Q$ increases, even for $\Lambda=0.1$, different values of $\zeta$ still exhibit clear distinctions in $|\mathrm{Im}(\omega)|$. This indicates that the influence of $\zeta$ remains visible at larger $Q$, consistent with the behavior observed for $\Lambda = 0.01$. Unlike the suppression effect of the angular quantum number $l$ on the quantum gravity corrections, increasing $\Lambda$ reduces numerical separations but does not eliminate the characteristic features associated with $\zeta$. Moreover, in other representative modes not shown here (e.g., $l=1$, overtone $n=1$, and purely imaginary modes), the characteristic effects induced by $\zeta$ remain evident, further confirming that the conclusions drawn from the $\Lambda=0.01$ data regarding the QNFs remain broadly applicable even at larger $\Lambda$.

Overall, the results for larger $\Lambda$ indicate that, although the cosmological constant reduces the numerical separation between different $\zeta$ cases, it does not alter the qualitative influence of quantum corrections on the QNF spectrum. Therefore, the main text presents only the results for the smaller value $\Lambda=0.01$, which both preserve the features of the dS background and more clearly illustrate the overall effect of the quantum parameter $\zeta$ on the QNFs.




\end{document}